\documentclass[aps,prb,twocolumn,superscriptaddress,showpacs,amsmath,amssymb]{revtex4-1}

\usepackage{graphicx}
\usepackage{dcolumn}
\usepackage{bm}
\usepackage[version=4]{mhchem}
\usepackage{pbox}
\usepackage{bibentry}
\usepackage{natbib}
\usepackage{color,soul}

\begin{document}


\title{\textit{Ab-initio} investigation of the thermodynamics of cation distribution and the electronic and magnetic structures in the \ce{LiMn2O4} spinel}


\author{David Santos-Carballal}
\email[]{SantosCarballalD@cardiff.ac.uk}
\affiliation{School of Chemistry, Cardiff University, Main Building, Park Place, Cardiff CF10 3AT, United Kingdom}
\affiliation{Materials Modelling Centre, School of Physical and Mineral Sciences, University of Limpopo, Private Bag x 1106, Sovenga, 0727, South Africa}

\author{Phuti E. Ngoepe}
\affiliation{Materials Modelling Centre, School of Physical and Mineral Sciences, University of Limpopo, Private Bag x 1106, Sovenga, 0727, South Africa}

\author{Nora H. de Leeuw}
\email[]{DeLeeuwN@cardiff.ac.uk}
\affiliation{School of Chemistry, Cardiff University, Main Building, Park Place, Cardiff CF10 3AT, United Kingdom}
\affiliation{Department of Earth Sciences, Utrecht University, Budapestlaan 4, 3584 CD Utrecht, The Netherlands}


\date{\today}

\begin{abstract}
The spinel-structured lithium manganese oxide (\ce{LiMn2O4}) is a material currently used as cathode for secondary lithium-ion batteries, but whose properties are not yet fully understood. Here, we report a computational investigation of the inversion thermodynamics and electronic behaviour of \ce{LiMn2O4} derived from spin-polarised density functional theory calculations with a Hubbard Hamiltonian and long-range dispersion corrections (DFT+\textit{U}\textendash D3). Based on the analysis of the configurational free energy, we have elucidated a partially inverse equilibrium cation distribution for the \ce{LiMn2O4} spinel. This equilibrium degree of inversion is rationalised in terms of the crystal field stabilisation effects and the difference between the size of the cations. We compare the atomic charges with the oxidation numbers for each degree of inversion. We found segregation of the Mn charge once these ions occupy the tetrahedral and octahedral sites of the spinel. We have obtained the atomic projections of the electronic band structure and density of states, showing that the normal \ce{LiMn2O4} has half-metallic properties, while the fully inverse spinel is an insulator. This material is in the ferrimagnetic state for the inverse and partially inverse cation arrangement. The optimised lattice and oxygen parameters, as well as the equilibrium degree of inversion, are in agreement with the available experimental data. The partially inverse equilibrium degree of inversion is important in the interpretation of the lithium ion migration and surface properties of the \ce{LiMn2O4} spinel.
\end{abstract}

\pacs{71.20.\textendash b, 71.15.Mb, 75.50.Gg, 82.60.\textendash s}

\maketitle


\section{INTRODUCTION}

The current academic and industrial interest in lithium manganese oxide \ce{LiMn2O4} spinel is based on its existing and future application as a cathode material for rechargeable (secondary) lithium-ion batteries.\cite{whittingham2004,*whittingham2008,*thackeray2012} LMO, short for the \ce{LiMn2O4} spinel, is considered the material of choice for the reduction cathodes due to its structural network of three-dimensional channels that allow the fast diffusion of lithium ions\cite{gummow1993,*thackeray1984a} during the battery charge and discharge, resulting in a relatively high rate for these processes.\cite{thackeray1983,tarascon1991} \ce{LiMn2O4} is an environmentally friendly compound and less toxic than the currently commercialised counterparts (\textit{e.g.}, \ce{LiCo2O4}).\cite{kim1997a} Moreover, the availability of the constituent elements and the thermal stability of this compound,\cite{cho1999} make it affordable and suitable for high power and sustainable energy applications.\cite{whittingham2004,*whittingham2008,*thackeray2012,armand2008,*arico2005,*tarascon2001}

Lithium is the smallest of the alkaline metals and as such it can be reversibly inserted at room temperature into the lattice of several spinels, forming compounds such as \ce{LiMn2O4}. The intercalated Li$_{1+y}$\ce{Mn2O4} system is cubic at $y=0$, but easily becomes tetragonal as $y$ starts to differ from zero.\cite{wickham1958,janovec1975,thackeray1983} \ce{LiMn2O4} is a mixed electronic/ionic conductor in which the mobile Li ion can diffuse throughout interconnected cavities within the \ce{Mn2O4} framework. The preferred method for preparing \ce{LiMn2O4} spinel is the solid-state reaction of a mixture in stoichiometric proportions of \ce{Li2CO3} and \ce{Mn2O3},\cite{wickham1958,strobel2004} or \ce{Mn2(CO3)3}\cite{strobel2004,ishizawa2014} in air at 650$^{\circ}$C.\cite{thackeray1983}

Bimetallic spinels of the type $CD_2\text{O}_4$ crystallise in the $Fd\bar{3}m$ space group (No. 227). These materials have an array of oxygen atoms forming a face-centred cubic (\textit{fcc}) sub-lattice with 1/8 of the tetrahedral interstitial sites and half of the octahedral sites occupied by the cations. Spinels with this stoichiometry can exist in a range of cation arrangements, whose extreme structures are the normal and inverse distribution. For the normal spinel, characterised by a unique structure of complete order, the $C$ cations are filling the tetrahedral positions ($A$) and the $D$ ions are restricted to the octahedral holes ($B$). However, the inverse structure has a range of configurations where half of the $D$ cations are occupying the tetrahedral positions and the other half along with the $C$ ions are distributed in the octahedral sub-lattice. The degree of inversion ($x$) quantifies the fraction of tetrahedral positions occupied by the $D$ cation when the spinel formula is more conveniently expressed as $(C_{1-x}D_x)_A(C_xD_{2-x})_B\text{O}_4$.\cite{barth1932} Each degree of inversion has a number of associated atomic configurations with reduced symmetry. However, the $Fd\bar{3}m$ symmetry is preserved if all the cation positions within each sub-lattice are effectively equivalent by considering that they are randomly distributed. Furthermore, the cubic \ce{LiMn2O4} structure suffers a first order phase transition at the Verwey-like temperature of $T_\text{V}=283.5$ K,\cite{tomeno2001} transforming either partially into a crystal with tetragonal space group $I4_1/amd$ (No. 141),\cite{yamada1995,*yamada1996} as supported by computational simulations,\cite{ouyang2009} or fully into the single orthorhombic symmetry $Fddd$ (No. 70).\cite{hayakawa1998,*rodriguez-carvajal1998,*oikawa1998} Both then convert further into the tetragonal phase,\cite{takada1999} due to the cooperative Jahn-Teller distortion of the Mn$^{3+}$ ions.\cite{yamaguchi1998}

The magnetic properties of \ce{LiMn2O4} have been widely reported in the literature,\cite{tomeno2001,yamada1995,*yamada1996,ouyang2009,hayakawa1998,*rodriguez-carvajal1998,*oikawa1998,takada1999,yamaguchi1998,wills1999,anderson1956,villain1979,reimers1991,*reimers1992,bhattacharya2013,karim2013,kim2015,aydinol1997a,*berg1999,vanderven2000,*koyama2003,shi2003,benedek2011,*tang2014,ouyang2010,jaber-ansari2015,kumar2014,nakayama2014,chukalkin2010} which has offered theoretical, computational and experimental insights into the various possible arrangements of the magnetic moments of the Mn atom. Although the magnetic properties of this spinel originate in the Mn unpaired electrons, the total magnetisation is given by the ground state configuration of the magnetic moments of these atoms under specific conditions. From a theoretical viewpoint, the high-spin Mn cations are antiferromagnetically coupled \textit{via} the super-exchange mechanism, where the closed-shell oxygen anions act as intermediaries. The super-exchange mechanism is most effective when the transition metal cations and the oxygen atoms are forming the angle $\angle$ Mn$-$O$-$Mn = 180$^{\circ}$.\cite{masquelier1996,*goodenough1955,*goodenough1955a,*goodenough1963,anderson1956} In the normal spinel, where all the magnetic carriers are in the octahedral positions and the angle $\angle$ Mn$_B-$O$-$Mn$_B$ $\approx$ 95$^{\circ}$, the weak and negative super-exchange interactions between them drives the antiparallel alignment of the magnetic moments of the Mn atoms, whereby the total magnetisation value of \ce{LiMn2O4} vanishes. However, these negative super-exchange interactions are stronger between the Mn ions populating different sub-lattices, as angle $\angle$ Mn$_A-$O$-$Mn$_B$ $\approx$ 120$^{\circ}$ is closer to 180$^{\circ}$ than within the same sub-lattice, leading to ferrimagnetism in the inverse \ce{LiMn2O4} spinel. The magnetic sub-lattice of the normal \ce{LiMn2O4} forms a pyrochlore network, where the Mn$_B$ atoms are located at the apices of corner-sharing tetrahedra.\cite{wills1999} Antiferromagnets with the pyrochlore atomic arrangement are geometrically frustrated systems,\cite{anderson1956,villain1979} as the atomic Mn$_B$ magnetic moments cannot simultaneously be aligned antiparallel to each other, forming a canted configuration in order to cancel the total magnetic moment.\cite{wills1999} This generates a ground state composed of a collection of structures with different spin configurations and degenerate energies.\cite{reimers1991,*reimers1992}

Most of the computational and experimental studies of the magnetic properties of \ce{LiMn2O4} have described the collinear short- and long-range magnetic orderings, leading to seemingly contradictory results as the non-collinear magnetic configurations were overlooked. For example, density functional theory simulations with a Hubbard Hamiltonian (DFT+\textit{U}) of the primitive cell of the normal \ce{LiMn2O4} suggest that the ferromagnetic ordering is 24 meV per formula unit (f.u.) lower in energy than the antiferromagnetic configuration,\cite{bhattacharya2013} highlighting the weak nature of the magnetic couplings between the Mn cations. Further computational studies using similar methods have shown that the antiferromagnetic configuration within the Mn chains along the [011] direction is more stable than the antiferromagnetic ordering where all the Mn atoms within one [001] plane are alternating their magnetic moments with the cations belonging to the neighbouring planes.\cite{ouyang2009} Meanwhile, computational studies of different antiferromagnetic orderings along the [011] direction have revealed that the symmetry-broken [$\uparrow\uparrow\downarrow\downarrow$] configuration is 28 meV/f.u. more stable than the [$\uparrow\downarrow\uparrow\downarrow$] configuration,\cite{karim2013} while the ferromagnetic configuration becomes degenerate with the [$\uparrow\uparrow\downarrow\downarrow$] arrangement if the symmetry is allowed to break.\cite{kim2015} Moreover, \textit{ab-initio} calculations using non spin-polarisation\cite{aydinol1997a,*berg1999} or the ferromagnetic\cite{vanderven2000,*koyama2003,bhattacharya2013} or antiferromagnetic\cite{shi2003} \ce{LiMn2O4} have predicted the average electrochemical open circuit voltage curve for the Li-intercalation reaction and defect formation energies in close agreement with experiments, showing that different magnetic arrangements have a negligible effect on the energetics of this spinel. Further calculations using first-principles methods have focused for simplicity on the ferromagnetic cubic cell\cite{benedek2011,*tang2014} or the antiferromagnetic orthorhombic cell,\cite{ouyang2010} correctly predicting a number of features for the low Miller index surfaces of the normal \ce{LiMn2O4}, regardless of the different magnetic orderings considered. DFT+\textit{U} calculations using the antiferromagnetic order along the [011] direction have elucidated the role of the single-layer graphene coatings in the mitigation of Mn loss throughout the (001) surface that has been observed experimentally.\cite{jaber-ansari2015} Although the above computational studies did not attempt to investigate intermediate inversion degrees on the \ce{LiMn2O4} bulk, Karim \textit{et al.} found that the surface Mn cations exchange sites with the sub-surface Li atoms, effectively reconstructing the (111) surface and suggesting a local partial inverse distribution for this spinel with the symmetry-broken antiferromagnetic [$\uparrow\uparrow\downarrow\downarrow$] ordering.\cite{karim2013} Following the seminal work by Karim and collaborators,\cite{karim2013} DFT+\textit{U} simulations using the partially inverse \ce{LiMn2O4}(111) surface slab have explained the mechanism for the catalytic decomposition of ethylene carbonate, one of the major components of battery electrolytes.\cite{kumar2014} They reproduced the electrochemical alternate current (AC) impedance measurements, as well as the experimentally determined chemical potential for the interfacial Li$^+$ exchange reaction, despite considering also the normal (001) and (011) surfaces.\cite{nakayama2014}

Experimental studies combining X-ray diffraction, neutron diffraction and susceptibility measurements have suggested the [$\uparrow\uparrow\downarrow\downarrow$] magnetic configuration along the [011] direction,\cite{tomeno2001} whereas more recent investigations have determined the long-range ordering vector of the Mn$_B$ magnetic moments in the normal \ce{LiMn2O4} spinel,\cite{wills1999,chukalkin2010} identifying the magnetic arrangement with the canted spins. As we are studying a number of intermediate degrees of inversion with the ferrimagnetic ordering (between the tetrahedral and octahedral positions), we have not attempted to simulate the normal spinel with the antiferromagnetic configuration (within the octahedral positions) as this will entail changing the magnetic arrangement as soon as the spinel starts to invert.

Here, we have used DFT+\textit{U} simulations to study how the degree of inversion modifies a number of properties in the room temperature cubic crystal structure of the \ce{LiMn2O4} spinel. We investigate the equilibrium cation arrangement using thermodynamic arguments and analyse the effect of the crystal field stabilisation effects and sizes of the cations. We also discuss the magnetic and electronic structure of \ce{LiMn2O4} for the extreme degrees of inversion.

 \begin{figure}
 \includegraphics[width=86mm,trim={0cm 1.0cm 0cm 1.0cm},clip]{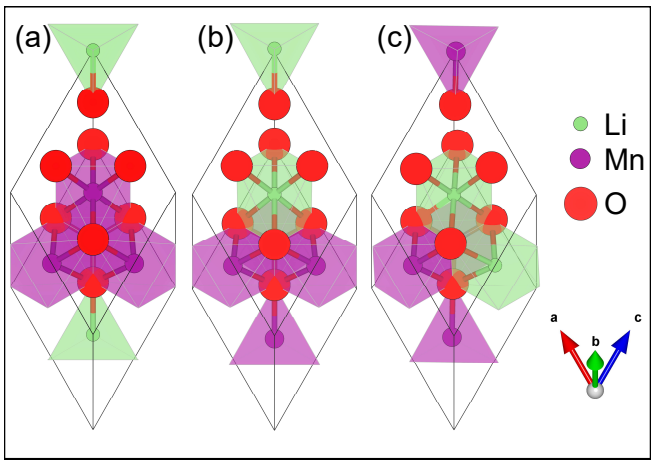}
 \caption{(Color online) Schematic representation of the (a) normal, (b) half-inverse and (c) fully-inverse rhombohedral primitive unit cell of the \ce{LiMn2O4} spinel.\label{fig:primitive_spinel}}
 \end{figure}

\section{COMPUTATIONAL METHODS}

\subsection{Calculation details}

The \ce{LiMn2O4} spinel structure was modelled using spin-polarised density functional theory (DFT) calculations as implemented in the Vienna \textit{Ab-initio} Simulation Package (\textsc{vasp}).\cite{kresse1993,*kresse1994,*kresse1996,*kresse1996a} All calculations were performed within the generalised gradient approximation (GGA) using the exchange-correlation functional of Perdew-Burke-Ernzerhof (PBE).\cite{Perdew1996,*Perdew1997} The core electrons and their interaction with the valence ones were described using the projector augmented-wave (PAW) method \cite{blochl1994} in the implementation of \textcite{kresse1999}. The frozen core consisted of up to and including the 3\textit{s} and 1\textit{s} orbitals for the Mn and O atom, respectively. For the Li atoms, all the electrons were treated as valence. The kinetic energy cutoff was fixed at 560 eV for the plane-wave basis set expansion of the Kohn-Sham (KS) valence states. A $\varGamma$-centred Monkhorst-Pack grid\cite{monkhorst1976} of $7\times7\times7$ $k$-points was used to carry out the integrations in the reciprocal space of the primitive unit cell. In order to improve the convergence of the Brillouin-Zone integrations, the electronic partial occupancies were determined using the Gaussian smearing, with a width for all calculations set at 0.05 eV. These smearing techniques are considered as a form of a finite temperature DFT, where the electronic free energy is the variational quantity.\cite{mermin1965} However, the tetrahedron method with Bl\"{o}chl corrections\cite{blochl1994a} was used for the calculation of the electronic and magnetic properties and to obtain very accurate total energies. The semi-empirical method of Grimme with the Becke-Johnson damping [D3-(BJ)] was also included in our calculations for the modelling of the long-range dispersion interactions.\cite{grimme2010a,*grimme2011} We have included the Van der Waals interactions in the simulation of the bulk phase of the typically ionic \ce{LiMn2O4} for consistency with future work. We expect to investigate the interaction of molecular adsorbates with the surfaces of this spinel, where dispersion effects are a major contribution to the energy and should therefore be included in the calculations.\cite{santos-carballal2014,*shields2015,*santos-carballal2016,*tafreshi2017,*roldan2017a,*dzade2017,*santos-carballal2017a} Geometry optimisations were conducted via the conjugate-gradient technique and were considered converged when the Hellmann-Feynman forces on all atoms dropped below 0.01 eV\r{A}$^{-1}$. All internal coordinates and lattice parameters were fully relaxed in the rhombohedral primitive unit cell, while ensuring that the conventional unit cell was always perfectly cubic.\cite{benedek2011,*tang2014}

A Hubbard correction\cite{anisimov1992} in the formulation of \textcite{dudarev1998} was applied to the Mn $3d$ orbitals to improve the description of the electron localisation. Our optimised effective parameter ($U_{\text{eff}}=4.0$ eV) is within those values reported in the literature for the simulation of \ce{LiMn2O4} spinel.\cite{bhattacharya2013,karim2013,ouyang2010,ouyang2009} We have developed the $U_{\text{eff}}$ value by fitting the position of the valence band maximum (VBM) and conduction band minimum (CVM) to those obtained from calculations using the screened hybrid functional of Heyd-Scuseria-Ernzerhof (HSE06).\cite{heyd2003,*heyd2006,*heyd2004,*heyd2004a,*heyd2005,*peralta2006,*krukau2006} For the fitting, we carried out single-point calculations with HSE06, using the structure with normal cation distribution and fully relaxed with the standard PBE functional. We then performed a set of full geometry optimisations followed by single-point calculations using PBE+$U$. In all the HSE06 simulations, we used the same settings as for the PBE functional.

The values of the initial magnetic moments were specified for the Mn atoms and were allowed to relax during the simulations. The initial moments were oriented parallel within each sub-lattice occupied by magnetism carriers and antiparallel to the other sub-lattice when it contained Mn atoms. Magnetic moments associated with different high- and low-spin electronic state combinations were tested for the Mn atoms. A Bader analysis of the atomic charges and spin moments was carried out by integrating these quantities within regions with zero flux of electronic density.\cite{henkelman2006,*sanville2007,*tang2009}

In order to determine the number of possible configurations that exist for \ce{LiMn2O4} supercells of different size, we have used the site-occupancy disorder (SOD) program.\cite{grau-crespo2007a}. This code produces the full configurational spectrum for each cell and identifies those symmetrically inequivalent configurations. Two configurations are considered symmetric if there is an isometric transformation derived from symmetry operators of the parent structure that relates these atom arrangements.

\subsection{Configurational free energy of inversion\label{sec:conf_free_ener_inv}}

For the analysis of the thermodynamics of inversion, we have treated the interchange of cations as a chemical equilibrium.\cite{navrotsky1967} The configurational free energy of inversion ($\varDelta F_{\text{config}}$) was determined in the usual way, by combining the configurational energy ($\varDelta E_{\text{config}}$) with the configurational entropy ($\varDelta S_{\text{config}}$) at a temperature $T$ according to the equation

\begin{equation}
\label{eq:config_free_ener}
\varDelta F_{\text{config}}=\varDelta E_{\text{config}}-T\varDelta S_{\text{config}}.
\end{equation}

We have assumed that the solid solution of the Li and Mn is ideal and have defined the complete ordered (normal) spinel as the standard state. Thus, $\varDelta E_{\text{config}}$ is calibrated with respect to the normal \ce{LiMn2O4} as $\varDelta E_{\text{config}}(x)=E_{\text{config}}(x)-E_{\text{config}}(x=0)$. The equation for $\varDelta S_{\text{config}}$ is expressed as the purely configurational random mixing of the two cations on the tetrahedral and octahedral sites as

\begin{eqnarray}
\label{eq:config_entrop}
\varDelta S_{\text{config}}=-R \left[x\ln x+(1-x)\ln(1-x)+x\ln\frac{x}{2}\right.\nonumber\\
\left.+(2-x)\ln\left(1-\frac{x}{2}\right)\right].
\end{eqnarray}

The entropy calculated via Eq. \ref{eq:config_entrop} takes the units from the ideal gas constant ($R$) and its value depends solely on the degree of inversion. $\varDelta S_{\text{config}}=0$ for the absolutely ordered normal spinel, in line with its definition as the standard state. $\varDelta S_{\text{config}}$ becomes increasingly positive as the Li and Mn cations are interchanged up to $x=2/3$, when the spinel reaches the complete disordered distribution. The configurational entropy decreases to the value of $2R\ln2$ for the fully inverse spinel. This methodology has previously been used successfully for the simulation of the inversion thermodynamics of a number of spinels.\cite{tielens2006,*palin2008,*seko2010,*ndione2014,seminovski2012,santos-carballal2015b}

\section{RESULTS AND DISCUSSION}

\subsection{Thermodynamics of the cation distribution in \ce{LiMn2O4}}

\subsubsection{Configurational spectrum}

We first investigate the distribution of the Li and Mn cations in the tetrahedral and octahedral positions of the rhombohedral primitive unit cell of the \ce{LiMn2O4} spinel. Our starting point is the fully ordered (normal) spinel structure with space group $Fd\bar{3}m$ (No. 227) determined by \textcite{strobel2004}, where the Li ions are distributed in the Wyckoff $8a$ tetrahedral positions and the Mn atoms in the $16d$ octahedral positions [see Fig. \ref{fig:primitive_spinel}(a)]. However, in order to simplify the notation used in this paper, we will refer to the tetrahedral and octahedral sub-lattices as ``$A$" and ``$B$", respectively.

Table \ref{tab:cation_configurations} lists the total number of cation configurations ($N$), together with the number of symmetrically inequivalent configurations ($M$) as function of the total number of formula units ($n_{\text{f.u.}}$), as determined using SOD. The total number of combinations of the two Li atoms over the $A$ and $B$ sites of the rhombohedral primitive unit cell ($n_{\text{f.u.}}=2$) is $6!/(2!\times4!)=15$, but only 3 of these are inequivalent. These 3 inequivalent configurations correspond to a degree of inversion of $x=0.0, 0.5 \text{ and } 1.0$, describing the lithium occupancy of 1, 1/2 and 0 on the $A$ sites (see Fig. \ref{fig:primitive_spinel}). To model additional degrees of inversion (\textit{i.e.}, $x=0.25 \text{ and } 0.75$) with integer occupancies of Li, we double the unit cell along one axis. Any axis can be chosen, given the symmetry of the rhombohedral cell, forming a $2\times1\times1$ supercell that contains four Li cations ($n_{\text{f.u.}}=4$). However, the simulation of this supercell becomes prohibitive as it is defined by 95 inequivalent configurations. Of course, larger supercells can be constructed to simulate other degrees of inversion, but this only increases the number of inequivalent configurations of our model. Thus, all the calculations in this work were carried out using the primitive unit cell with $n_{\text{f.u.}}=2$ unless otherwise stated. This model has been found sufficient to appropriately describe the spinel properties for these degrees of inversion.\cite{wei2001,*walsh2007,fritsch2010,*fritsch2011b,seminovski2012}

\begin{table}
\caption{Total number of cation configurations ($N$) and number of symmetrically inequivalent configurations ($M$) as a function of the total number of formula units ($n_{\text{f.u.}}$) of \ce{LiMn2O4} per simulation cell.\label{tab:cation_configurations}}
\begin{ruledtabular}
\begin{tabular}{ddd}
n_{\text{f.u.}} & N & M\\
\colrule
2 & 15 & 3\\
4 & 495 & 95\\
6 & 18,564 & 1,831\\
8 & 735,471 & 4,222\\
\end{tabular}
\end{ruledtabular}
\end{table}

\subsubsection{Equilibrium configurational structures}

Table \ref{tab:lattice_properties} summarises the relaxed unit cell ($a$) and oxygen ($u$) parameter for the three simulated degrees of inversion ($x=0.0, 0.5 \text{ and } 1.0$). Clearly, our calculations somewhat overestimate the experimental lattice parameter for any degree of inversion. However, we found that the optimised oxygen parameter is in better agreement with experiments than the lattice parameter, with only a reported difference in the third decimal place. In general, the fully inverse spinel exhibits the smallest deviation from experiment for both structural parameters, with $a$ overestimated by 0.036 \r{A} and $u$ underestimated by 0.002. We should note that the oxygen parameter for all degrees of inversion is above the ideal value of 0.250 expected for the perfect closest-packing arrangement of the oxygen atoms in the spinels. This deviation indicates that the oxygen atoms move in the [111] direction as the interstitial tetrahedral and octahedral sites are filled by cations of specific sizes.

\begin{table}
\caption{Summary of the unit cell lattice ($a$) and oxygen ($u$) parameter of the \ce{LiMn2O4} spinel reported from experiments (Ref. \onlinecite{strobel2004}) and calculated for the degree of inversion $x=0.0, 0.5 \text{ and } 1.0$. Note that the unit cell origin is defined at the centre of symmetry.\cite{brabers1995}\label{tab:lattice_properties}}
\begin{ruledtabular}
\begin{tabular}{lcccc}
 & Experimental & $x=0.0$ & $x=0.5$ & $x=1.0$\\
\colrule
$a$ (\r{A}) & 8.247 & 8.350 & 8.287 & 8.283\\
$u$ & 0.263 & 0.268 & 0.266 & 0.261\\
\end{tabular}
\end{ruledtabular}
\end{table}

\subsubsection{Inversion thermodynamics and equilibrium degree of inversion}

\begin{figure}
\includegraphics[width=86mm]{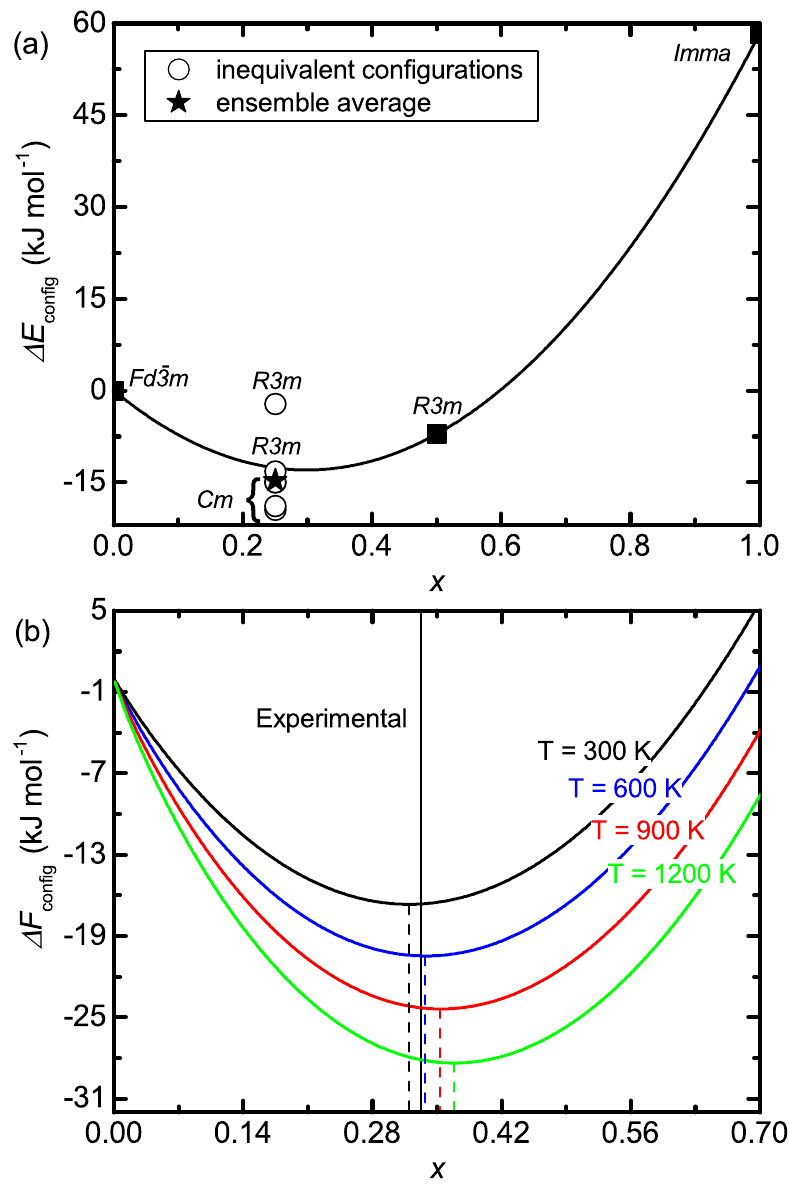}
\caption{(Color online) (a) Configurational inversion energy ($\varDelta E_{\text{config}}$) and (b) configurational inversion free energy ($\varDelta F_{\text{config}}$) both per formula unit and as a function of the degree of inversion ($x$) for the \ce{LiMn2O4} spinel. The open circles correspond to the inequivalent configurations with $x=0.25$ in a $2\times1\times1$ supercell. The star represents the ensemble average of these configurations. The space group of the inequivalent configurations is also shown. The dashed lines mark the equilibrium degree of inversion at various temperatures. The solid vertical line indicates the experimental value of the equilibrium degree of inversion, from Ref. \onlinecite{chukalkin2010}.\label{fig:inversion_energies}}
\end{figure}

The calculated values for the inversion configurational energy per formula unit ($\varDelta E_{\text{config}}$) indicate that the half-inverse \ce{LiMn2O4} spinel, with space group $R3m$ (No. 160), is the most stable structure [see Fig. \ref{fig:inversion_energies}(a)]. We also found that the normal spinel, with the $Fd\bar{3}m$ symmetry, is $\sim60 \text{ kJ mol}^{-1}$ more favourable than the cation arrangement with degree of inversion $x=1.0$ and space group $Imma$ (No. 74). This result is in good agreement with previous DFT calculations, where the antiferromagnetic character of \ce{LiMn2O4} was not considered (for the degrees of inversion $x>0.0$),\cite{bhattacharya2013} suggesting that the trend in the inversion thermodynamics does not depend on different spin configurations. Interestingly, the above artificial lowering of the symmetry is the result of swapping the site of the Li for the Mn atom in the primitive unit cell of the spinel. Therefore, any symmetry lowering is the result of the choice of the simulation cell, as the periodic boundary conditions dictate a particular cation arrangement with perfect long-range order. Obviously, the use of incommensurate cells circumvents this problem by allowing a quasi-random configurations of cations, but this comes at the cost of excessive simulation times.\cite{fritsch2010,*fritsch2011b} In order to predict the configurational energy for any value of the degree of inversion, we have have used a quadratic equation. This type of fitting, initially empirically proposed by Kriessman and Harrison\cite{kriessman1956} and later theoretically confirmed by \textcite{o'neill1983}, has been used for the modelling of different spinels, providing results in close agreement with experiments.\cite{seminovski2012,santos-carballal2015b} Figure \ref{fig:inversion_energies}(a) shows that the configurational inversion energy is negative for the degree of inversion $0.0 < x < 0.6$, becoming positive for the rest of the range of $x$.

To validate the equation for the above fitting, we have calculated the configurational energy for the $2\times1\times1$ supercell with the degree of inversion $x=0.25$. For this purpose, we have generated the six inequivalent configurations, whose energies are spread from $-2.16$ to $-13.21 \text{ kJ mol}^{-1}$ [see Fig. \ref{fig:inversion_energies}(a)]. The two structures with the highest $R3m$ symmetry have the highest value of the configurational energy. On the other hand, the cation arrangements with the lower symmetry, \textit{i.e.}, with space group $Cm$ (No. 8) have larger degeneracy and are clustered in a narrower range of inversion energy. Thus, the average configuration energy of the ensemble for the supercell with degree of inversion $x=0.25$ is only around $2 \text{ kJ mol}^{-1}$ lower in energy than the quadratic interpolation, supporting the model based on the results from the primitive unit cell.

We have obtained the configurational free energy of inversion ($\varDelta F_{\text{config}}$) from $\varDelta E_{\text{config}}$ and the configurational inversion entropy ($\varDelta S_{\text{config}}$), following the procedure explained in section \ref{sec:conf_free_ener_inv}. We have estimated the free energy of inversion between 300 and 1200 K, \textit{i.e.}, room temperature and the typical synthesis temperature, respectively [see Fig. \ref{fig:inversion_energies}(b)]. At the temperatures under consideration, the \ce{LiMn2O4} spinel is predicted as partially inverse with $0.32<x<0.37$ under equilibrium conditions as a result of the minimum of the inversion energy at $x=0.3$. To the best of our knowledge, this equilibrium degree of inversion has not been found in previous computational works as the half-inverse \ce{LiMn2O4} spinel has thus far not been modelled. However, our equilibrium degree of inversion is in excellent agreement with the value of $x=0.333$ reported by Chukalkin and co-workers after irradiating samples of this material with fast neutrons at 340 K.\cite{chukalkin2010}

\subsection{Electronic structure of \ce{LiMn2O4}}

Although the thermodynamics of inversion provide evidence of the equilibrium cation distribution in spinels, they do not offer an explanation at the atomic or electronic level. The equilibrium degree of inversion has traditionally been rationalised in terms of the interplay between the relative sizes of the cations and the crystal field stabilisation energy for these ions in the tetrahedral and octahedral coordination environments. As such, the value of any observable property derived from the electronic and magnetic structure of the \ce{LiMn2O4} spinel will depend on the equilibrium cation configuration. For intermediate degrees of inversion, the expected value of any such properties can be interpolated from the normal and inverse spinel. In this section we examine the effect of the sizes of the cations and the crystal field stabilisation energy on the equilibrium cation arrangement in the \ce{LiMn2O4} spinel, followed by the analysis of the electronic and magnetic properties of the completely normal ($x=0.0$) and fully inverse ($x=1.0$) \ce{LiMn2O4} and the deduced behaviour for the spinel with partial equilibrium degrees of inversion.

\subsubsection{Size of the cations and crystal field stabilisation energy}

For the analysis of the sizes of the cations, we have assumed that all ions are rigid spheres lying in contact. This model allows us to define the ratio between the tetrahedral ($R_{A}$) and octahedral ($R_{B}$) bond distances exclusively as a function of the oxygen parameter $u$. Taking into account that the tetrahedral sites are larger than the octahedral ones for $u>0.2625$,\cite{hill1979} which is the case for the normal ($x=0.0$) and half-inverse ($x=0.5$) spinel (see Table \ref{tab:lattice_properties}), we expect that $R_{B}<R_{A}$ for the stable \ce{LiMn2O4}.

We have used the Shannon effective radii, which depend on the coordination number, electronic spin and oxidation number,\cite{shannon1976} as they provide a good indication of the bond distance of the ion in question. The ${\text{Li}_{A}}^{+}$ has a smaller radius than ${\text{Mn}_{B}}^{3+}$ with a high-spin electronic distribution, resulting in an inverse cation arrangement, yet the normal spinel is favoured when considering that ${\text{Mn}_{B}}^{4+}$ ions are smaller than ${\text{Li}_{A}}^{+}$. The opposed tendency of the $\text{Mn}^{3+}$ and $\text{Mn}^{4+}$ ions when filling the tetrahedral and octahedral positions explains qualitatively the intermediate degree of inversion of the mixed valence \ce{LiMn2O4} spinel.

\begin{figure}
 \includegraphics[width=86mm,trim={0cm 2.5cm 0cm 2.5cm},clip]{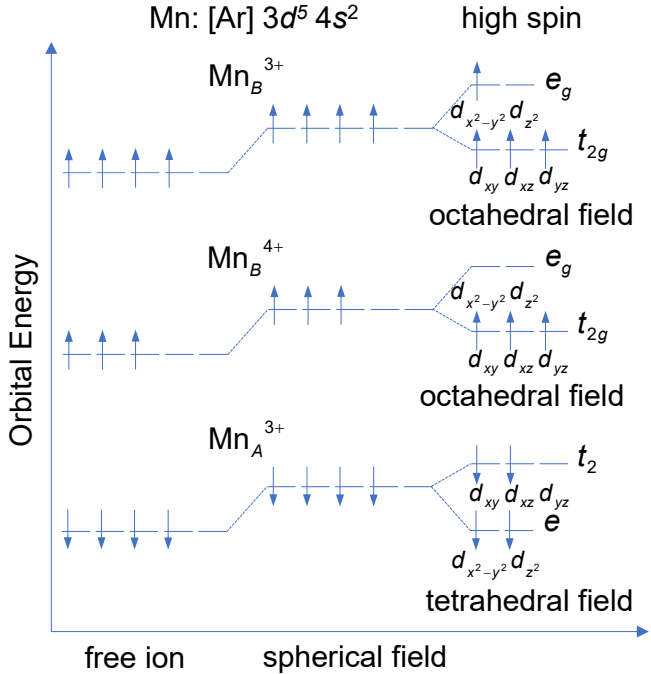}
 \caption{(Color online) Scheme of the crystal field splitting for the Mn 3$d$ electrons in high spin tetrahedral and octahedral coordination environments. The perturbation of the Mn 3$d$ electrons in a spherical field is also represented.\label{fig:crystal_field_splitting}}
 \end{figure}

The crystal field stabilisation is an important factor to consider when comparing the feasibility of the Li and Mn atoms to occupy the two types of cation positions within the spinel. The crystal field stabilisation energy measures the splitting of the open shell \textit{d} levels in transition metals placed in a field of anions. Fig. \ref{fig:crystal_field_splitting} shows the crystal field splitting and the expected electronic occupation of the Mn 3$d$ orbitals for the formal 3+ and 4+ oxidation states in the tetrahedral and octahedral coordination environments. Thus, the crystal field stabilisation energy for cations in a tetrahedral and octahedral environment of oxygen ions\cite{mcclure1957,dunitz1957} provides an estimate of the relative stability of the normal and inverse \ce{LiMn2O4} spinel. In our case, we have only used the difference between the tetrahedral and octahedral stabilisation energies (octahedral site preference energy [OSPE]) for the Mn ion as Li is not a transition metal. The OSPE value of 95.4 kJ mol$^{-1}$ for Mn$^{3+}$ and the lack of preference for the Li$^{+}$ ion, clearly favours the normal \ce{LiMn2O4} spinel. On the other hand, the quantitative OSPE value for Mn$^{4+}$ has not been experimentally determined due to the strong additional Jahn-Teller stabilisation of the tetrahedral geometry. However, it is qualitatively known that $d^{3}$ ions such as Mn$^{4+}$ have the $t_{2g}$ level half-full and a larger preference for the octahedral coordination environments than ions with any other electronic configuration (see Fig. \ref{fig:crystal_field_splitting}),\cite{dunitz1957} which leads to the conclusion that Mn$^{4+}$ ions will tend to remain in the octahedral positions once the \ce{LiMn2O4} spinel becomes (partially) inverse.

\begin{table*}
\caption{Atomic charges ($q$), atomic spin moments ($m_{\text{s}}$) and total spin magnetisation of saturation ($M_{\text{S}}$) per formula unit (f.u.) for the normal ($x=0.0$) and inverse ($x=1.0$) \ce{LiMn2O4} spinel. All properties were calculated by means of a Bader analysis.\label{tab:atomic_properties}}
\begin{ruledtabular}
\begin{tabular}{dddddddddddd}
 &\multicolumn{5}{c}{$q\text{ }(\text{e atom}^{-1})$} & \multicolumn{5}{c}{$m_{\text{s}}\text{ }(\mu_{\text{B}} \text{ atom}^{-1})$} & \multicolumn{1}{c}{$M_{\text{S}}$} \\\cline{2-6}\cline{7-11}
\multicolumn{1}{c}{$x$} & \multicolumn{1}{c}{$\text{Li}_A$} & \multicolumn{1}{c}{$\text{Mn}_A$} & \multicolumn{1}{c}{$\text{Li}_B$} & \multicolumn{1}{c}{$\text{Mn}_B$} & \multicolumn{1}{c}{O} & \multicolumn{1}{c}{$\text{Li}_A$} & \multicolumn{1}{c}{$\text{Mn}_A$} & \multicolumn{1}{c}{$\text{Li}_B$} & \multicolumn{1}{c}{$\text{Mn}_B$} & \multicolumn{1}{c}{O} & \multicolumn{1}{c}{$\text{ }(\mu_{\text{B}} \text{ f.u.}^{-1})$} \\
\colrule
0.0 & 0.88 & \multicolumn{1}{c}{$-$} & \multicolumn{1}{c}{$-$} & 1.78 & -1.11 & 0.01 & \multicolumn{1}{c}{$-$} & \multicolumn{1}{c}{$-$} & 3.66 & -0.08 & 7.00 \\
1.0 & \multicolumn{1}{c}{$-$} & 1.67 & 0.88 & 1.88 & -1.11 & \multicolumn{1}{c}{$-$} & -3.89 & 0.00 & 2.91 & 0.00 & -1.00 \\
\end{tabular}
\end{ruledtabular}
\end{table*}

\subsubsection{Atomic charges and spin moments\label{sec:atom_charg_spin_mom}}

Table \ref{tab:atomic_properties} summarises the charge ($q$) transferred between the neutral atoms upon the formation of \ce{LiMn2O4}. All the calculated ionic charges are systematically underestimated for any degree of inversion, which is normal for Bader charges.\cite{roldan2013,santos-carballal2015b} The charge of the monovalent Li is 12\% smaller than its oxidation state, while for the Mn in the normal spinel it is $\sim$49\% below the average effective oxidation state of 3.5+. The lack of charge separation within the Mn sublattice in the normal spinel is consistent with the unique symmetry of the octahedral positions they are filling in the room-temperature cubic crystal structure. However, it should be noted that Mn charge segregation has been found if the cubic symmetry is allowed to break to obtain the crystal structures reported below the Verwey-like temperature.\cite{zhou2004,*hoang2014} Our simulations show that the Li and O charges are not sensitive to the cation arrangement. Moreover, the Mn charge of the atoms occupying the octahedral position $B$ in the normal spinel is intermediate between those in the tetrahedral $A$ and octahedral $B$ positions of the inverse spinel. The relative charges of the Mn atoms in the tetrahedral and octahedral sites of the inverse spinel suggests that $q_A<q_{B}$.

Table \ref{tab:atomic_properties} also shows the atomic spin moments per atom ($m_s$) for the \ce{LiMn2O4} spinel. As expected, the Li and O atoms are non-magnetic or have very low magnetic moments in the normal and inverse spinel, respectively. When $x=0.0$, the spin moment of $\text{Mn}_B$ is overestimated by 0.16 $\mu_{\text{B}} \text{ atom}^{-1}$ compared with the one expected for the electronic distribution of an equal mixture of the Mn${_B}^{3+}$: $t_{2g\uparrow}^3e_{g\uparrow}^1$ and Mn${_B}^{4+}$: $t_{2g\uparrow}^3$. In the inverse spinel, we found that our calculations underestimate slightly the expected value for the spin moments of the Mn$_A$ and Mn$_{B}$ ions. Our results also indicate that the Mn atom with the largest spin moment has high-spin electronic distribution and is located in the tetrahedral position, in agreement with the charge analysis. The underestimated atomic spin moments can be interpreted as the result of the itinerant electron magnetism.\cite{moriya1984} It is well known that the $3d$ itinerant electrons interact strongly with one another and are not localised on atoms but move from atom to atom within the crystal.

 \begin{figure*}
 \includegraphics[width=172mm]{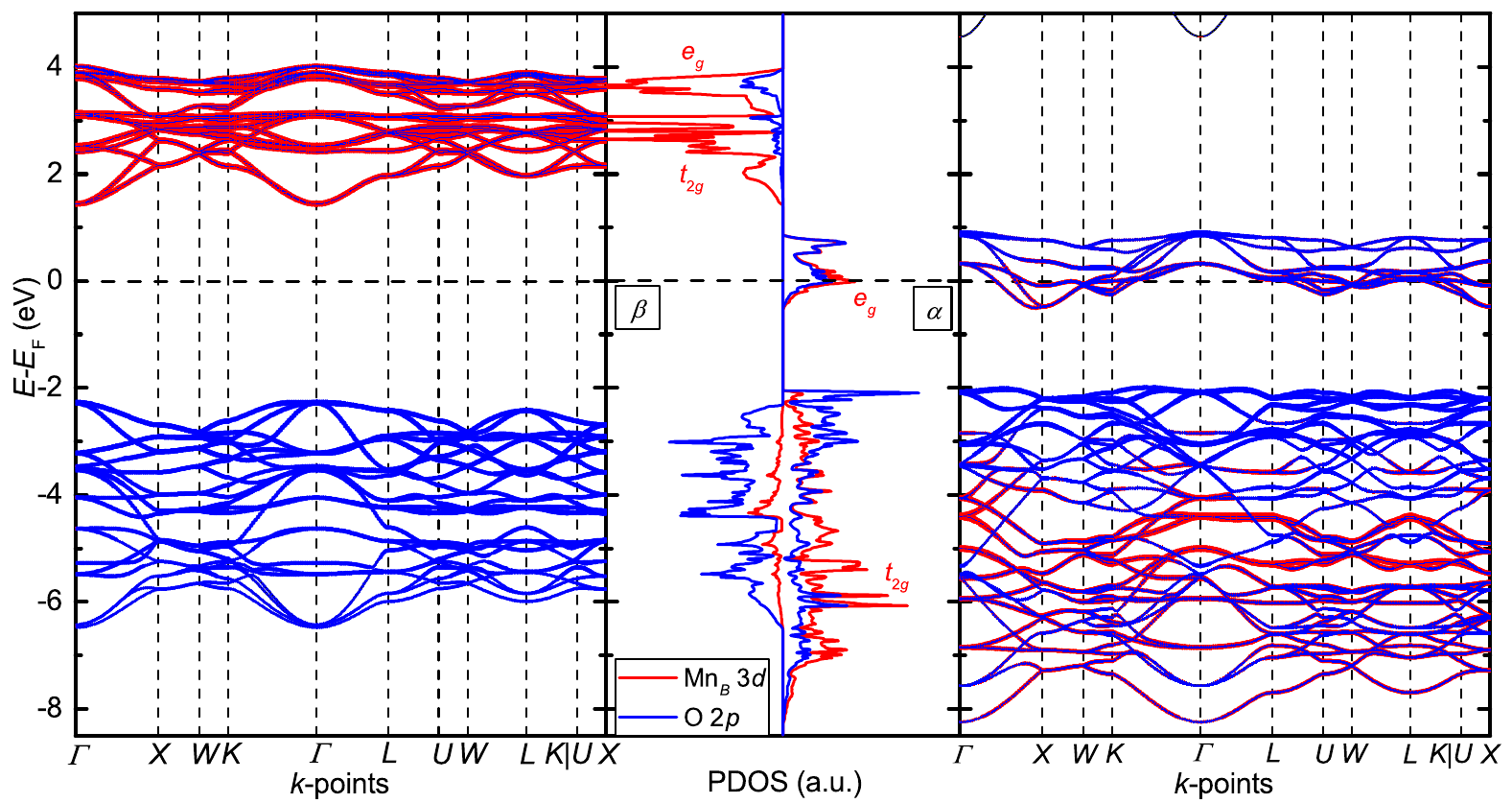}
 \caption{(Color online) Atomic projections of the spin decomposed electronic band structure (left and right panels) and total density of states (PDOS) (middle panel) for \ce{LiMn2O4} with degree of inversion $x=0.0$. $\alpha$ and $\beta$ stand for the majority and minority channel of the spins, respectively.\label{fig:dos_normal}}
 \end{figure*}

Figure \ref{fig:crystal_field_splitting} displays a representation of the electronic configuration of the Mn ions with formal oxidation numbers 3+ and 4+ when placed in octahedral and tetrahedral fields of anions. As discussed above, the charges and magnetic moments are the same for all the Mn$_B$ atoms, rendering them equivalent in the normal spinel, whose crystal symmetry ($Fd\bar{3}m$ space group) is therefore not affected. The indistinguishable Mn$_B$ ions can be understood on the basis that one electron is hopping from the Mn$^{3+}$ to the Mn$^{4+}$, in agreement with the good electronic conduction properties of this material.\cite{massarotti1997} Although the Mn$_B$ charges are underestimated (1.78 e$^-$/atom), the prediction of the spin moments (3.66 $\mu_{\text{B}}$/atom) is in much better agreement, with 3.5 unpaired electrons for the average charge of Mn$^{3+}$ and Mn$^{4+}$ atoms in a 1:1 ratio. Upon inversion, the crystal symmetry is reduced to the $Imma$ space group, as half of the Mn ions occupy the tetrahedral positions. Here, we found segregation of the Mn charges for this cation configuration. The oxidation numbers inferred from the Bader charges are again under-estimated, but the spin moments for the Mn$_A$ (3.89 $\mu_{\text{B}}$/atom) and the Mn$_B$ (2.91 $\mu_{\text{B}}$/atom) are in even closer agreement with 4 and 3 unpaired electrons, respectively, than in the normal spinel. The preference of the Mn$^{4+}$ ions to remain in the octahedral positions deduced from the OSPE for the inverse spinel is also in agreement with the spin moments of the Mn$_A$ and Mn$_B$ atoms.

In order to gain a more complete characterisation of the \ce{LiMn2O4} properties, we have calculated the magnetisation of saturation (total spin magnetisation) $M_S$ following the so-called N\'{e}el model.\cite{neel1948} This property, which quantifies the largest magnetisation per formula unit that a material can attain under an increasing magnetic field, is obtained as the sum of the atomic spin moments of the tetrahedral, octahedral and oxygen sub-lattices. We found that the \ce{LiMn2O4} spinel is ferrimagnetic for all degrees of inversion $x>0.0$. In this way, as the spinel gains disorder, the total magnetisation decreases. The $M_S$ values for the two extreme degrees of inversion can be interpolated to infer the magnetisation of saturation of the spinel with the equilibrium degree of inversion of $x=0.3$. Assuming a linear dependence of the total spin magnetisation with respect to $x$, we can propose that $M_S=4.6$ $\mu_{\text{B}} \text{ f.u.}^{-1}$ for the conditions of thermodynamic equilibrium.
 \begin{figure*}
 \includegraphics[width=172mm]{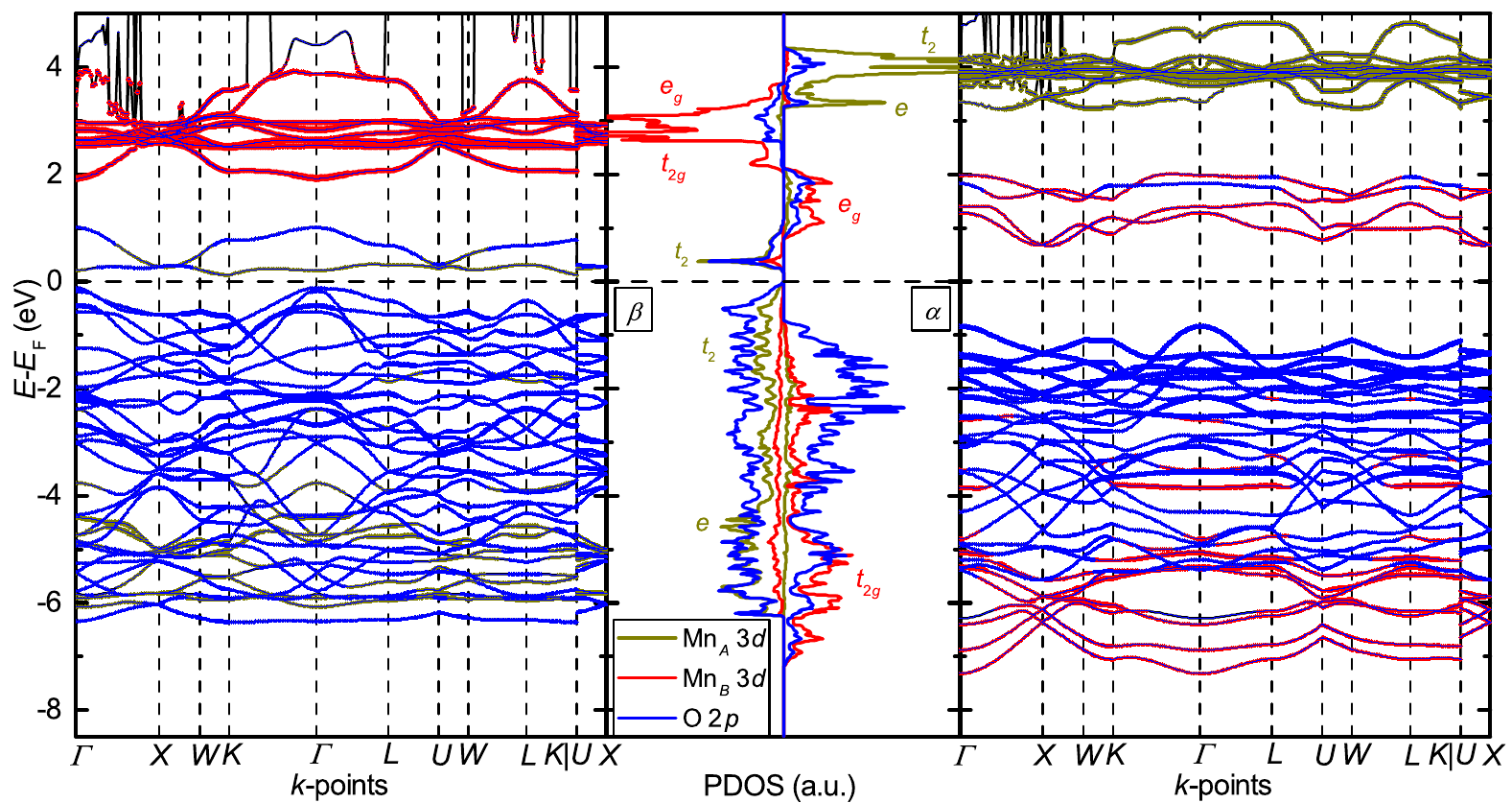}
 \caption{(Color online) Atomic projections of the spin decomposed electronic band structure (left and right panels) and total density of states (PDOS) (middle panel) for \ce{LiMn2O4} with degree of inversion $x=1.0$. $\alpha$ and $\beta$ stand for the majority and minority channel of the spins, respectively.\label{fig:dos_inverse}}
 \end{figure*}

\subsubsection{Density of states}

The density of states (DOS) in Fig. \ref{fig:dos_normal} shows that the normal spinel is half-metallic. At the Fermi energy ($E_F$), the spin-up ($\alpha$ channel) partially occupied $e_g$ level of the Mn${_B}^{3+}$ ions appears strongly hybridised with the O $2p$ orbitals, while the minority spin channel ($\beta$) is insulator. The other valence bands of the Mn$_B$ ions ($t_{2g}$) in the majority channel of the spins dominate the bottom of the valence band at $-6$.0 eV. The unoccupied $t_{2g}$ and $e_g$ levels of Mn$_B$ appear at 2.8 and 3.7 eV in the majority channel of the spins, with the latter one weakly hybridized with the O $2p$ orbitals. The contribution from the occupied O $2p$ orbitals to the DOS is spread between $-8.5$ and $-2.0$ eV in the $\alpha$ spin channel and between $-6.5$ and $-2.2$ eV in the $\beta$ channel. The empty levels of the anion are highly localised and coupled with the Mn$_B$ levels.

The inversion of all the Li cations to half of the octahedral positions in \ce{LiMn2O4} generates two types of $B$ positions, the Mn$_B$ containing 50\% of Mn and the Li$_B$ with all the Li (see Fig. \ref{fig:dos_inverse}). With this cation distribution, the spinel becomes an insulator with the smallest band-gap predicted in the minority channel of the spins. As discussed in subsection \ref{sec:atom_charg_spin_mom}, the Mn ions suffer charge segregation as a result of the \ce{LiMn2O4} inversion. The occupied $t_{2g}$ band of the highest charged Mn$_B$ ion is located approximately in the same position as in the normal spinel, although less intense and strongly hybridised with the O $2p$ orbitals. The $t_{2g}$ and $e_g$ levels in the minority spin channel are merged altogether at 2.8 eV, and the unoccupied $e_g$ band in the $\alpha$ channel is shifted to 1.5 eV with respect to Mn$_B$ in the normal spinel. The $e$ bands and part of the $t_2$ bands of the Mn$_A$ ion in the $\beta$ spin channel are occupied and are highly hybridised with the O $2p$ orbitals. The rest of the Mn$_A$ levels are unoccupied and therefore in the conduction band. The O $2p$ valence band runs from $-7.2$ and $-0.8$ eV in the majority spin channel, while the band edges are shifted $\sim0.8$ eV in the $\beta$ channel. As in the normal spinel, the unoccupied O orbitals have relatively low intensity and are contained within the cation bands.

We have found that the Li ion contribution to the DOS in the $-8.5$ to 5.0 eV range is negligible for any degree of inversion. Thus, this cation does not directly influence any of the electro-magnetic properties of the \ce{LiMn2O4} spinel. Moreover, the half-metallic and insulating properties of the normal and fully inverse \ce{LiMn2O4} spinels, respectively, were confirmed by the integer values calculated for the total spin magnetisation (see Table \ref{tab:atomic_properties} and Figs. \ref{fig:dos_normal} and \ref{fig:dos_inverse}). More details regarding the relationship between the total number of electrons and the conducting properties of a given material can be found elsewhere.\cite{santos-carballal2015b}

\subsubsection{Electronic band structure}

The electronic band structure depicted in Figs. \ref{fig:dos_normal} and \ref{fig:dos_inverse} shows that $\varGamma$, $K$ and $X$ are the three high symmetry points, which contribute to the valence and conduction band extrema. Therefore, we focus our attention on the direct and relevant indirect transitions for each degree of inversion and spin channel where a band gap exists.

For the $\beta$ channel of the normal spinel, the VBM is dominated by the O $2p$ orbitals, while the CBM is mostly derived from the Mn$_B$ $t_{2g}$ levels. The calculated $\varGamma$$-$$\varGamma$ transition energy is 3.7 eV for this spin channel and degree of inversion. We do not calculate the energy of any indirect transition as the VBM and CBM are at the $\varGamma$ point for the minority channel of the normal \ce{LiMn2O4}.

For the degree of inversion $x=1.0$, there is a band gap opening in the majority spin channel. The top edge of this band gap is composed of a mixture of states resulting from both the Mn$_B$ $e_g$ states and the O $2p$ orbitals. Our DFT simulations also show that the indirect $\varGamma$ to $X$ transition is 0.6 eV lower in energy than the direct transitions at $\varGamma$ and $X$ that are close in energy. Upon inversion, the composition of the VBM and CBM of the $\beta$ spin channel remain largely unchanged with respect to the normal spinel. However, the electronic band gap suffers a reduction and the lowest energy transition is indirect and occurs between $\varGamma$ and $K$ (0.1 eV). We did not find evidence of the Mn$_A$ cations playing any direct role in the electronic properties of the inverse spinel.

\section{SUMMARY AND CONCLUSIONS}

In this study we have reported the electronic and magnetic structure along with the equilibrium cation distribution of the \ce{LiMn2O4} spinel. We have modelled all the cation arrangements of four degrees of inversion by using the primitive unit cell and a $2\times1\times1$ supercell. The equilibrium degree of inversion determined between room temperature and the firing temperature shows that Li and Mn have approximately a 7:3 ratio preference for the tetrahedral sites in agreement with the available experimental data. The calculated lattice and oxygen parameters for the different degrees of inversion compare well with those obtained from X-ray crystallographic data. The relative size of the cations explain qualitatively the intermediate equilibrium degree of inversion found for \ce{LiMn2O4}, but no conclusion can be drawn from the crystal field stabilisation effects

As it is commonly found, all atomic charges are underestimated with respect to the oxidation numbers for the normal and fully inverse spinel. The Mn atoms experience a charge segregation upon the spinel inversion, with those holding the lower charge located in the tetrahedral positions $A$. The Li and O charges do not change for any of the simulated degrees of inversion. The largest total spin magnetisation was calculated for the degree of inversion $x=0.0$ as all the magnetism carriers are aligned parallel in the octahedral position $B$. We have predicted the magnetisation of saturation for the calculated equilibrium degree of inversion of \ce{LiMn2O4}. We found that Li and Mn$_A$ ions do not affect directly the electronic properties for any of the extreme degrees of inversion. The normal spinel is described as a half-metal, while the inverse one is as an insulator. The band gap in the $\beta$ channel of the normal \ce{LiMn2O4} is due to a direct electronic transition at the $\varGamma$ point. However, the band gap that opens up in the two spin channels of the inverse spinel are associated with an indirect electronic transition with the $\varGamma$ point always at the edge of the conduction band.

Future work will involve Monte Carlo simulations of larger supercells to sample effectively the entire configurational spectrum and determine the lowest energy configuration with the degree of inversion $x=0.3$. This will allow us to study the interaction of the Li-ion battery electrolyte components with the surfaces of the partially inverse \ce{LiMn2O4} spinel.

\begin{acknowledgments}
D.S.\textendash C. is grateful to the Department of Science and Technology (DST) and the National Research Foundation (NRF) of South Africa for the provision of a Postdoctoral Fellowship for Early Career Researchers from the United Kingdom. We acknowledge the Engineering \& Physical Sciences Research Council (EPSRC grants No. EP/K009567/2 and No. EP/K016288/1) for funding. We acknowledge the use of the Centre for High Performance Computing (CHPC) facility of South Africa in the completion of this work. Via our membership of the U.K.\textquoteright s HEC Materials Chemistry Consortium, which is funded by EPSRC (EP/L000202), this work used ARCHER, the U.K. National Supercomputing Service (http://www.archer.ac.uk). This work was performed using the computational facilities of the Advanced Research Computing @ Cardiff (ARCCA) Division, Cardiff University. The authors also acknowledge the use of HPC Wales, and associated support services, in the completion of this work. All data created during this research is openly available from the Cardiff University\textquoteright s Research Portal at http://dx.doi.org/10.17035/d.2017.0032205417.
\end{acknowledgments}

\bibliography{library}

\begin{thebibliography}{105}%
\makeatletter
\providecommand \@ifxundefined [1]{%
 \@ifx{#1\undefined}
}%
\providecommand \@ifnum [1]{%
 \ifnum #1\expandafter \@firstoftwo
 \else \expandafter \@secondoftwo
 \fi
}%
\providecommand \@ifx [1]{%
 \ifx #1\expandafter \@firstoftwo
 \else \expandafter \@secondoftwo
 \fi
}%
\providecommand \natexlab [1]{#1}%
\providecommand \enquote  [1]{``#1''}%
\providecommand \bibnamefont  [1]{#1}%
\providecommand \bibfnamefont [1]{#1}%
\providecommand \citenamefont [1]{#1}%
\providecommand \href@noop [0]{\@secondoftwo}%
\providecommand \href [0]{\begingroup \@sanitize@url \@href}%
\providecommand \@href[1]{\@@startlink{#1}\@@href}%
\providecommand \@@href[1]{\endgroup#1\@@endlink}%
\providecommand \@sanitize@url [0]{\catcode `\\12\catcode `\$12\catcode
  `\&12\catcode `\#12\catcode `\^12\catcode `\_12\catcode `\%12\relax}%
\providecommand \@@startlink[1]{}%
\providecommand \@@endlink[0]{}%
\providecommand \url  [0]{\begingroup\@sanitize@url \@url }%
\providecommand \@url [1]{\endgroup\@href {#1}{\urlprefix }}%
\providecommand \urlprefix  [0]{URL }%
\providecommand \Eprint [0]{\href }%
\providecommand \doibase [0]{http://dx.doi.org/}%
\providecommand \selectlanguage [0]{\@gobble}%
\providecommand \bibinfo  [0]{\@secondoftwo}%
\providecommand \bibfield  [0]{\@secondoftwo}%
\providecommand \translation [1]{[#1]}%
\providecommand \BibitemOpen [0]{}%
\providecommand \bibitemStop [0]{}%
\providecommand \bibitemNoStop [0]{.\EOS\space}%
\providecommand \EOS [0]{\spacefactor3000\relax}%
\providecommand \BibitemShut  [1]{\csname bibitem#1\endcsname}%
\let\auto@bib@innerbib\@empty
\bibitem [{\citenamefont {Whittingham}(2004)}]{whittingham2004}%
  \BibitemOpen
  \bibfield  {author} {\bibinfo {author} {\bibfnamefont {M.~S.}\ \bibnamefont
  {Whittingham}},\ }\href {\doibase 10.1021/cr020731c} {\bibfield  {journal}
  {\bibinfo  {journal} {Chem. Rev.}\ }\textbf {\bibinfo {volume} {104}},\
  \bibinfo {pages} {4271} (\bibinfo {year} {2004})}\BibitemShut {NoStop}%
\bibitem [{\citenamefont {Whittingham}(2008)}]{whittingham2008}%
  \BibitemOpen
  \bibfield  {author} {\bibinfo {author} {\bibfnamefont {M.~S.}\ \bibnamefont
  {Whittingham}},\ }\href {\doibase 10.1557/mrs2008.82} {\bibfield  {journal}
  {\bibinfo  {journal} {MRS Bull.}\ }\textbf {\bibinfo {volume} {33}},\
  \bibinfo {pages} {411} (\bibinfo {year} {2008})}\BibitemShut {NoStop}%
\bibitem [{\citenamefont {Thackeray}\ \emph {et~al.}(2012)\citenamefont
  {Thackeray}, \citenamefont {Wolverton},\ and\ \citenamefont
  {Isaacs}}]{thackeray2012}%
  \BibitemOpen
  \bibfield  {author} {\bibinfo {author} {\bibfnamefont {M.~M.}\ \bibnamefont
  {Thackeray}}, \bibinfo {author} {\bibfnamefont {C.}~\bibnamefont
  {Wolverton}}, \ and\ \bibinfo {author} {\bibfnamefont {E.~D.}\ \bibnamefont
  {Isaacs}},\ }\href {\doibase 10.1039/c2ee21892e} {\bibfield  {journal}
  {\bibinfo  {journal} {Energy Environ. Sci.}\ }\textbf {\bibinfo {volume}
  {5}},\ \bibinfo {pages} {7854} (\bibinfo {year} {2012})}\BibitemShut
  {NoStop}%
\bibitem [{\citenamefont {Gummow}\ \emph {et~al.}(1993)\citenamefont {Gummow},
  \citenamefont {Liles},\ and\ \citenamefont {Thackeray}}]{gummow1993}%
  \BibitemOpen
  \bibfield  {author} {\bibinfo {author} {\bibfnamefont {R.~J.}\ \bibnamefont
  {Gummow}}, \bibinfo {author} {\bibfnamefont {D.~C.}\ \bibnamefont {Liles}}, \
  and\ \bibinfo {author} {\bibfnamefont {M.~M.}\ \bibnamefont {Thackeray}},\
  }\href {\doibase 10.1016/0025-5408(93)90172-A} {\bibfield  {journal}
  {\bibinfo  {journal} {Mater. Res. Bull.}\ }\textbf {\bibinfo {volume} {28}},\
  \bibinfo {pages} {1249} (\bibinfo {year} {1993})}\BibitemShut {NoStop}%
\bibitem [{\citenamefont {Thackeray}\ \emph {et~al.}(1984)\citenamefont
  {Thackeray}, \citenamefont {Johnson}, \citenamefont {de~Picciotto},
  \citenamefont {Bruce},\ and\ \citenamefont {Goodenough}}]{thackeray1984a}%
  \BibitemOpen
  \bibfield  {author} {\bibinfo {author} {\bibfnamefont {M.~M.}\ \bibnamefont
  {Thackeray}}, \bibinfo {author} {\bibfnamefont {P.~J.}\ \bibnamefont
  {Johnson}}, \bibinfo {author} {\bibfnamefont {L.~A.}\ \bibnamefont
  {de~Picciotto}}, \bibinfo {author} {\bibfnamefont {P.~G.}\ \bibnamefont
  {Bruce}}, \ and\ \bibinfo {author} {\bibfnamefont {J.~B.}\ \bibnamefont
  {Goodenough}},\ }\href {\doibase 10.1016/0025-5408(84)90088-6} {\bibfield
  {journal} {\bibinfo  {journal} {Mater. Res. Bull.}\ }\textbf {\bibinfo
  {volume} {19}},\ \bibinfo {pages} {179} (\bibinfo {year} {1984})}\BibitemShut
  {NoStop}%
\bibitem [{\citenamefont {Thackeray}\ \emph {et~al.}(1983)\citenamefont
  {Thackeray}, \citenamefont {David}, \citenamefont {Bruce},\ and\
  \citenamefont {Goodenough}}]{thackeray1983}%
  \BibitemOpen
  \bibfield  {author} {\bibinfo {author} {\bibfnamefont {M.~M.}\ \bibnamefont
  {Thackeray}}, \bibinfo {author} {\bibfnamefont {W.~I.~F.}\ \bibnamefont
  {David}}, \bibinfo {author} {\bibfnamefont {P.~G.}\ \bibnamefont {Bruce}}, \
  and\ \bibinfo {author} {\bibfnamefont {J.~B.}\ \bibnamefont {Goodenough}},\
  }\href {\doibase 10.1016/0025-5408(83)90138-1} {\bibfield  {journal}
  {\bibinfo  {journal} {Mater. Res. Bull.}\ }\textbf {\bibinfo {volume} {18}},\
  \bibinfo {pages} {461} (\bibinfo {year} {1983})}\BibitemShut {NoStop}%
\bibitem [{\citenamefont {Tarascon}\ \emph {et~al.}(1991)\citenamefont
  {Tarascon}, \citenamefont {Wang}, \citenamefont {Shokoohi}, \citenamefont
  {McKinnon},\ and\ \citenamefont {Colson}}]{tarascon1991}%
  \BibitemOpen
  \bibfield  {author} {\bibinfo {author} {\bibfnamefont {J.~M.}\ \bibnamefont
  {Tarascon}}, \bibinfo {author} {\bibfnamefont {E.}~\bibnamefont {Wang}},
  \bibinfo {author} {\bibfnamefont {F.~K.}\ \bibnamefont {Shokoohi}}, \bibinfo
  {author} {\bibfnamefont {W.~R.}\ \bibnamefont {McKinnon}}, \ and\ \bibinfo
  {author} {\bibfnamefont {S.}~\bibnamefont {Colson}},\ }\href {\doibase
  10.1149/1.2085330} {\bibfield  {journal} {\bibinfo  {journal} {J.
  Electrochem. Soc.}\ }\textbf {\bibinfo {volume} {138}},\ \bibinfo {pages}
  {2859} (\bibinfo {year} {1991})}\BibitemShut {NoStop}%
\bibitem [{\citenamefont {Kim}\ and\ \citenamefont
  {Manthiram}(1997)}]{kim1997a}%
  \BibitemOpen
  \bibfield  {author} {\bibinfo {author} {\bibfnamefont {J.}~\bibnamefont
  {Kim}}\ and\ \bibinfo {author} {\bibfnamefont {A.}~\bibnamefont
  {Manthiram}},\ }\href {\doibase 10.1038/36812} {\bibfield  {journal}
  {\bibinfo  {journal} {Nature}\ }\textbf {\bibinfo {volume} {390}},\ \bibinfo
  {pages} {265} (\bibinfo {year} {1997})}\BibitemShut {NoStop}%
\bibitem [{\citenamefont {Cho}\ and\ \citenamefont {Kim}(1999)}]{cho1999}%
  \BibitemOpen
  \bibfield  {author} {\bibinfo {author} {\bibfnamefont {J.}~\bibnamefont
  {Cho}}\ and\ \bibinfo {author} {\bibfnamefont {G.}~\bibnamefont {Kim}},\
  }\href {\doibase 10.1149/1.1390802} {\bibfield  {journal} {\bibinfo
  {journal} {Electrochem. Solid-State Lett.}\ }\textbf {\bibinfo {volume}
  {2}},\ \bibinfo {pages} {253} (\bibinfo {year} {1999})}\BibitemShut {NoStop}%
\bibitem [{\citenamefont {Armand}\ and\ \citenamefont
  {Tarascon}(2008)}]{armand2008}%
  \BibitemOpen
  \bibfield  {author} {\bibinfo {author} {\bibfnamefont {M.}~\bibnamefont
  {Armand}}\ and\ \bibinfo {author} {\bibfnamefont {J.-M.}\ \bibnamefont
  {Tarascon}},\ }\href {\doibase 10.1038/451652a} {\bibfield  {journal}
  {\bibinfo  {journal} {Nature}\ }\textbf {\bibinfo {volume} {451}},\ \bibinfo
  {pages} {652} (\bibinfo {year} {2008})}\BibitemShut {NoStop}%
\bibitem [{\citenamefont {Aric{\`{o}}}\ \emph {et~al.}(2005)\citenamefont
  {Aric{\`{o}}}, \citenamefont {Bruce}, \citenamefont {Scrosati}, \citenamefont
  {Tarascon},\ and\ \citenamefont {van Schalkwijk}}]{arico2005}%
  \BibitemOpen
  \bibfield  {author} {\bibinfo {author} {\bibfnamefont {A.~S.}\ \bibnamefont
  {Aric{\`{o}}}}, \bibinfo {author} {\bibfnamefont {P.}~\bibnamefont {Bruce}},
  \bibinfo {author} {\bibfnamefont {B.}~\bibnamefont {Scrosati}}, \bibinfo
  {author} {\bibfnamefont {J.-M.}\ \bibnamefont {Tarascon}}, \ and\ \bibinfo
  {author} {\bibfnamefont {W.}~\bibnamefont {van Schalkwijk}},\ }\href
  {\doibase 10.1038/nmat1368} {\bibfield  {journal} {\bibinfo  {journal} {Nat.
  Mater.}\ }\textbf {\bibinfo {volume} {4}},\ \bibinfo {pages} {366} (\bibinfo
  {year} {2005})}\BibitemShut {NoStop}%
\bibitem [{\citenamefont {Tarascon}\ and\ \citenamefont
  {Armand}(2001)}]{tarascon2001}%
  \BibitemOpen
  \bibfield  {author} {\bibinfo {author} {\bibfnamefont {J.-M.}\ \bibnamefont
  {Tarascon}}\ and\ \bibinfo {author} {\bibfnamefont {M.}~\bibnamefont
  {Armand}},\ }\href {\doibase 10.1038/35104644} {\bibfield  {journal}
  {\bibinfo  {journal} {Nature}\ }\textbf {\bibinfo {volume} {414}},\ \bibinfo
  {pages} {359} (\bibinfo {year} {2001})}\BibitemShut {NoStop}%
\bibitem [{\citenamefont {Wickham}\ and\ \citenamefont
  {Croft}(1958)}]{wickham1958}%
  \BibitemOpen
  \bibfield  {author} {\bibinfo {author} {\bibfnamefont {D.~G.}\ \bibnamefont
  {Wickham}}\ and\ \bibinfo {author} {\bibfnamefont {W.~J.}\ \bibnamefont
  {Croft}},\ }\href {\doibase 10.1016/0022-3697(58)90285-3} {\bibfield
  {journal} {\bibinfo  {journal} {J. Phys. Chem. Solids}\ }\textbf {\bibinfo
  {volume} {7}},\ \bibinfo {pages} {351} (\bibinfo {year} {1958})}\BibitemShut
  {NoStop}%
\bibitem [{\citenamefont {Janovec}\ \emph {et~al.}(1975)\citenamefont
  {Janovec}, \citenamefont {Dvoř{\'{a}}k},\ and\ \citenamefont
  {Petzelt}}]{janovec1975}%
  \BibitemOpen
  \bibfield  {author} {\bibinfo {author} {\bibfnamefont {V.}~\bibnamefont
  {Janovec}}, \bibinfo {author} {\bibfnamefont {V.}~\bibnamefont
  {Dvoř{\'{a}}k}}, \ and\ \bibinfo {author} {\bibfnamefont {J.}~\bibnamefont
  {Petzelt}},\ }\href {\doibase 10.1007/BF01587561} {\bibfield  {journal}
  {\bibinfo  {journal} {Czechoslov. J. Phys. B}\ }\textbf {\bibinfo {volume}
  {25}},\ \bibinfo {pages} {1362} (\bibinfo {year} {1975})}\BibitemShut
  {NoStop}%
\bibitem [{\citenamefont {Strobel}\ \emph {et~al.}(2004)\citenamefont
  {Strobel}, \citenamefont {Rousse}, \citenamefont {Ibarra-Palos},\ and\
  \citenamefont {Masquelier}}]{strobel2004}%
  \BibitemOpen
  \bibfield  {author} {\bibinfo {author} {\bibfnamefont {P.}~\bibnamefont
  {Strobel}}, \bibinfo {author} {\bibfnamefont {G.}~\bibnamefont {Rousse}},
  \bibinfo {author} {\bibfnamefont {A.}~\bibnamefont {Ibarra-Palos}}, \ and\
  \bibinfo {author} {\bibfnamefont {C.}~\bibnamefont {Masquelier}},\ }\href
  {\doibase 10.1016/S0022-4596(03)00189-0} {\bibfield  {journal} {\bibinfo
  {journal} {J. Solid State Chem.}\ }\textbf {\bibinfo {volume} {177}},\
  \bibinfo {pages} {1} (\bibinfo {year} {2004})}\BibitemShut {NoStop}%
\bibitem [{\citenamefont {Ishizawa}\ \emph {et~al.}(2014)\citenamefont
  {Ishizawa}, \citenamefont {Tateishi}, \citenamefont {Oishi},\ and\
  \citenamefont {Kishimoto}}]{ishizawa2014}%
  \BibitemOpen
  \bibfield  {author} {\bibinfo {author} {\bibfnamefont {N.}~\bibnamefont
  {Ishizawa}}, \bibinfo {author} {\bibfnamefont {K.}~\bibnamefont {Tateishi}},
  \bibinfo {author} {\bibfnamefont {S.}~\bibnamefont {Oishi}}, \ and\ \bibinfo
  {author} {\bibfnamefont {S.}~\bibnamefont {Kishimoto}},\ }\href {\doibase
  10.2138/am.2014.4840} {\bibfield  {journal} {\bibinfo  {journal} {Am.
  Mineral.}\ }\textbf {\bibinfo {volume} {99}},\ \bibinfo {pages} {1528}
  (\bibinfo {year} {2014})}\BibitemShut {NoStop}%
\bibitem [{\citenamefont {Barth}\ and\ \citenamefont
  {Posnjak}(1932)}]{barth1932}%
  \BibitemOpen
  \bibfield  {author} {\bibinfo {author} {\bibfnamefont {T.~F.~W.}\
  \bibnamefont {Barth}}\ and\ \bibinfo {author} {\bibfnamefont
  {E.}~\bibnamefont {Posnjak}},\ }\href {\doibase 10.1524/zkri.1932.82.1.325}
  {\bibfield  {journal} {\bibinfo  {journal} {Zeitschrift f{\"{u}}r Krist. -
  Cryst. Mater.}\ }\textbf {\bibinfo {volume} {82}},\ \bibinfo {pages} {325}
  (\bibinfo {year} {1932})}\BibitemShut {NoStop}%
\bibitem [{\citenamefont {Tomeno}\ \emph {et~al.}(2001)\citenamefont {Tomeno},
  \citenamefont {Kasuya},\ and\ \citenamefont {Tsunoda}}]{tomeno2001}%
  \BibitemOpen
  \bibfield  {author} {\bibinfo {author} {\bibfnamefont {I.}~\bibnamefont
  {Tomeno}}, \bibinfo {author} {\bibfnamefont {Y.}~\bibnamefont {Kasuya}}, \
  and\ \bibinfo {author} {\bibfnamefont {Y.}~\bibnamefont {Tsunoda}},\ }\href
  {\doibase 10.1103/PhysRevB.64.094422} {\bibfield  {journal} {\bibinfo
  {journal} {Phys. Rev. B}\ }\textbf {\bibinfo {volume} {64}},\ \bibinfo
  {pages} {094422} (\bibinfo {year} {2001})}\BibitemShut {NoStop}%
\bibitem [{\citenamefont {Yamada}\ and\ \citenamefont
  {Tanaka}(1995)}]{yamada1995}%
  \BibitemOpen
  \bibfield  {author} {\bibinfo {author} {\bibfnamefont {A.}~\bibnamefont
  {Yamada}}\ and\ \bibinfo {author} {\bibfnamefont {M.}~\bibnamefont
  {Tanaka}},\ }\href {\doibase 10.1016/0025-5408(95)00048-8} {\bibfield
  {journal} {\bibinfo  {journal} {Mater. Res. Bull.}\ }\textbf {\bibinfo
  {volume} {30}},\ \bibinfo {pages} {715} (\bibinfo {year} {1995})}\BibitemShut
  {NoStop}%
\bibitem [{\citenamefont {Yamada}(1996)}]{yamada1996}%
  \BibitemOpen
  \bibfield  {author} {\bibinfo {author} {\bibfnamefont {A.}~\bibnamefont
  {Yamada}},\ }\href {\doibase 10.1006/jssc.1996.0097} {\bibfield  {journal}
  {\bibinfo  {journal} {J. Solid State Chem.}\ }\textbf {\bibinfo {volume}
  {122}},\ \bibinfo {pages} {160} (\bibinfo {year} {1996})}\BibitemShut
  {NoStop}%
\bibitem [{\citenamefont {Ouyang}\ \emph {et~al.}(2009)\citenamefont {Ouyang},
  \citenamefont {Shi},\ and\ \citenamefont {Lei}}]{ouyang2009}%
  \BibitemOpen
  \bibfield  {author} {\bibinfo {author} {\bibfnamefont {C.~Y.}\ \bibnamefont
  {Ouyang}}, \bibinfo {author} {\bibfnamefont {S.~Q.}\ \bibnamefont {Shi}}, \
  and\ \bibinfo {author} {\bibfnamefont {M.~S.}\ \bibnamefont {Lei}},\ }\href
  {\doibase 10.1016/j.jallcom.2008.06.123} {\bibfield  {journal} {\bibinfo
  {journal} {J. Alloys Compd.}\ }\textbf {\bibinfo {volume} {474}},\ \bibinfo
  {pages} {370} (\bibinfo {year} {2009})}\BibitemShut {NoStop}%
\bibitem [{\citenamefont {Hayakawa}\ \emph {et~al.}(1998)\citenamefont
  {Hayakawa}, \citenamefont {Takada}, \citenamefont {Enoki},\ and\
  \citenamefont {Akiba}}]{hayakawa1998}%
  \BibitemOpen
  \bibfield  {author} {\bibinfo {author} {\bibfnamefont {H.}~\bibnamefont
  {Hayakawa}}, \bibinfo {author} {\bibfnamefont {T.}~\bibnamefont {Takada}},
  \bibinfo {author} {\bibfnamefont {H.}~\bibnamefont {Enoki}}, \ and\ \bibinfo
  {author} {\bibfnamefont {E.}~\bibnamefont {Akiba}},\ }\href {\doibase
  10.1023/A:1006682304966} {\bibfield  {journal} {\bibinfo  {journal} {J.
  Mater. Sci. Lett.}\ }\textbf {\bibinfo {volume} {17}},\ \bibinfo {pages}
  {811} (\bibinfo {year} {1998})}\BibitemShut {NoStop}%
\bibitem [{\citenamefont {Rodr{\'{i}}guez-Carvajal}\ \emph
  {et~al.}(1998)\citenamefont {Rodr{\'{i}}guez-Carvajal}, \citenamefont
  {Rousse}, \citenamefont {Masquelier},\ and\ \citenamefont
  {Hervieu}}]{rodriguez-carvajal1998}%
  \BibitemOpen
  \bibfield  {author} {\bibinfo {author} {\bibfnamefont {J.}~\bibnamefont
  {Rodr{\'{i}}guez-Carvajal}}, \bibinfo {author} {\bibfnamefont
  {G.}~\bibnamefont {Rousse}}, \bibinfo {author} {\bibfnamefont
  {C.}~\bibnamefont {Masquelier}}, \ and\ \bibinfo {author} {\bibfnamefont
  {M.}~\bibnamefont {Hervieu}},\ }\href {\doibase 10.1103/PhysRevLett.81.4660}
  {\bibfield  {journal} {\bibinfo  {journal} {Phys. Rev. Lett.}\ }\textbf
  {\bibinfo {volume} {81}},\ \bibinfo {pages} {4660} (\bibinfo {year}
  {1998})}\BibitemShut {NoStop}%
\bibitem [{\citenamefont {Oikawa}(1998)}]{oikawa1998}%
  \BibitemOpen
  \bibfield  {author} {\bibinfo {author} {\bibfnamefont {K.}~\bibnamefont
  {Oikawa}},\ }\href {\doibase 10.1016/S0167-2738(98)00073-3} {\bibfield
  {journal} {\bibinfo  {journal} {Solid State Ionics}\ }\textbf {\bibinfo
  {volume} {109}},\ \bibinfo {pages} {35} (\bibinfo {year} {1998})}\BibitemShut
  {NoStop}%
\bibitem [{\citenamefont {Takada}\ \emph {et~al.}(1999)\citenamefont {Takada},
  \citenamefont {Hayakawa}, \citenamefont {Enoki}, \citenamefont {Akiba},
  \citenamefont {Slegr}, \citenamefont {Davidson},\ and\ \citenamefont
  {Murray}}]{takada1999}%
  \BibitemOpen
  \bibfield  {author} {\bibinfo {author} {\bibfnamefont {T.}~\bibnamefont
  {Takada}}, \bibinfo {author} {\bibfnamefont {H.}~\bibnamefont {Hayakawa}},
  \bibinfo {author} {\bibfnamefont {H.}~\bibnamefont {Enoki}}, \bibinfo
  {author} {\bibfnamefont {E.}~\bibnamefont {Akiba}}, \bibinfo {author}
  {\bibfnamefont {H.}~\bibnamefont {Slegr}}, \bibinfo {author} {\bibfnamefont
  {I.}~\bibnamefont {Davidson}}, \ and\ \bibinfo {author} {\bibfnamefont
  {J.}~\bibnamefont {Murray}},\ }\href {\doibase 10.1016/S0378-7753(98)00225-0}
  {\bibfield  {journal} {\bibinfo  {journal} {J. Power Sources}\ }\textbf
  {\bibinfo {volume} {81-82}},\ \bibinfo {pages} {505} (\bibinfo {year}
  {1999})}\BibitemShut {NoStop}%
\bibitem [{\citenamefont {Yamaguchi}\ \emph {et~al.}(1998)\citenamefont
  {Yamaguchi}, \citenamefont {Yamada},\ and\ \citenamefont
  {Uwe}}]{yamaguchi1998}%
  \BibitemOpen
  \bibfield  {author} {\bibinfo {author} {\bibfnamefont {H.}~\bibnamefont
  {Yamaguchi}}, \bibinfo {author} {\bibfnamefont {A.}~\bibnamefont {Yamada}}, \
  and\ \bibinfo {author} {\bibfnamefont {H.}~\bibnamefont {Uwe}},\ }\href
  {\doibase 10.1103/PhysRevB.58.8} {\bibfield  {journal} {\bibinfo  {journal}
  {Phys. Rev. B}\ }\textbf {\bibinfo {volume} {58}},\ \bibinfo {pages} {8}
  (\bibinfo {year} {1998})}\BibitemShut {NoStop}%
\bibitem [{\citenamefont {Wills}\ \emph {et~al.}(1999)\citenamefont {Wills},
  \citenamefont {Raju},\ and\ \citenamefont {Greedan}}]{wills1999}%
  \BibitemOpen
  \bibfield  {author} {\bibinfo {author} {\bibfnamefont {A.~S.}\ \bibnamefont
  {Wills}}, \bibinfo {author} {\bibfnamefont {N.~P.}\ \bibnamefont {Raju}}, \
  and\ \bibinfo {author} {\bibfnamefont {J.~E.}\ \bibnamefont {Greedan}},\
  }\href {\doibase 10.1021/cm981041l} {\bibfield  {journal} {\bibinfo
  {journal} {Chem. Mater.}\ }\textbf {\bibinfo {volume} {11}},\ \bibinfo
  {pages} {1510} (\bibinfo {year} {1999})}\BibitemShut {NoStop}%
\bibitem [{\citenamefont {Anderson}(1956)}]{anderson1956}%
  \BibitemOpen
  \bibfield  {author} {\bibinfo {author} {\bibfnamefont {P.~W.}\ \bibnamefont
  {Anderson}},\ }\href {\doibase 10.1103/PhysRev.102.1008} {\bibfield
  {journal} {\bibinfo  {journal} {Phys. Rev.}\ }\textbf {\bibinfo {volume}
  {102}},\ \bibinfo {pages} {1008} (\bibinfo {year} {1956})}\BibitemShut
  {NoStop}%
\bibitem [{\citenamefont {Villain}(1979)}]{villain1979}%
  \BibitemOpen
  \bibfield  {author} {\bibinfo {author} {\bibfnamefont {J.}~\bibnamefont
  {Villain}},\ }\href {\doibase 10.1007/BF01325811} {\bibfield  {journal}
  {\bibinfo  {journal} {Zeitschrift f{\"{u}}r Phys. B Condens. Matter Quanta}\
  }\textbf {\bibinfo {volume} {33}},\ \bibinfo {pages} {31} (\bibinfo {year}
  {1979})}\BibitemShut {NoStop}%
\bibitem [{\citenamefont {Reimers}\ \emph {et~al.}(1991)\citenamefont
  {Reimers}, \citenamefont {Greedan}, \citenamefont {Kremer}, \citenamefont
  {Gmelin},\ and\ \citenamefont {Subramanian}}]{reimers1991}%
  \BibitemOpen
  \bibfield  {author} {\bibinfo {author} {\bibfnamefont {J.~N.}\ \bibnamefont
  {Reimers}}, \bibinfo {author} {\bibfnamefont {J.~E.}\ \bibnamefont
  {Greedan}}, \bibinfo {author} {\bibfnamefont {R.~K.}\ \bibnamefont {Kremer}},
  \bibinfo {author} {\bibfnamefont {E.}~\bibnamefont {Gmelin}}, \ and\ \bibinfo
  {author} {\bibfnamefont {M.~A.}\ \bibnamefont {Subramanian}},\ }\href
  {\doibase 10.1103/PhysRevB.43.3387} {\bibfield  {journal} {\bibinfo
  {journal} {Phys. Rev. B}\ }\textbf {\bibinfo {volume} {43}},\ \bibinfo
  {pages} {3387} (\bibinfo {year} {1991})}\BibitemShut {NoStop}%
\bibitem [{\citenamefont {Reimers}(1992)}]{reimers1992}%
  \BibitemOpen
  \bibfield  {author} {\bibinfo {author} {\bibfnamefont {J.~N.}\ \bibnamefont
  {Reimers}},\ }\href {\doibase 10.1103/PhysRevB.45.7287} {\bibfield  {journal}
  {\bibinfo  {journal} {Phys. Rev. B}\ }\textbf {\bibinfo {volume} {45}},\
  \bibinfo {pages} {7287} (\bibinfo {year} {1992})}\BibitemShut {NoStop}%
\bibitem [{\citenamefont {Bhattacharya}\ and\ \citenamefont
  {Wolverton}(2013)}]{bhattacharya2013}%
  \BibitemOpen
  \bibfield  {author} {\bibinfo {author} {\bibfnamefont {J.}~\bibnamefont
  {Bhattacharya}}\ and\ \bibinfo {author} {\bibfnamefont {C.}~\bibnamefont
  {Wolverton}},\ }\href {\doibase 10.1039/c3cp50910a} {\bibfield  {journal}
  {\bibinfo  {journal} {Phys. Chem. Chem. Phys.}\ }\textbf {\bibinfo {volume}
  {15}},\ \bibinfo {pages} {6486} (\bibinfo {year} {2013})}\BibitemShut
  {NoStop}%
\bibitem [{\citenamefont {Karim}\ \emph {et~al.}(2013)\citenamefont {Karim},
  \citenamefont {Fosse},\ and\ \citenamefont {Persson}}]{karim2013}%
  \BibitemOpen
  \bibfield  {author} {\bibinfo {author} {\bibfnamefont {A.}~\bibnamefont
  {Karim}}, \bibinfo {author} {\bibfnamefont {S.}~\bibnamefont {Fosse}}, \ and\
  \bibinfo {author} {\bibfnamefont {K.~A.}\ \bibnamefont {Persson}},\ }\href
  {\doibase 10.1103/PhysRevB.87.075322} {\bibfield  {journal} {\bibinfo
  {journal} {Phys. Rev. B}\ }\textbf {\bibinfo {volume} {87}},\ \bibinfo
  {pages} {075322} (\bibinfo {year} {2013})}\BibitemShut {NoStop}%
\bibitem [{\citenamefont {Kim}\ \emph {et~al.}(2015)\citenamefont {Kim},
  \citenamefont {Aykol},\ and\ \citenamefont {Wolverton}}]{kim2015}%
  \BibitemOpen
  \bibfield  {author} {\bibinfo {author} {\bibfnamefont {S.}~\bibnamefont
  {Kim}}, \bibinfo {author} {\bibfnamefont {M.}~\bibnamefont {Aykol}}, \ and\
  \bibinfo {author} {\bibfnamefont {C.}~\bibnamefont {Wolverton}},\ }\href
  {\doibase 10.1103/PhysRevB.92.115411} {\bibfield  {journal} {\bibinfo
  {journal} {Phys. Rev. B}\ }\textbf {\bibinfo {volume} {92}},\ \bibinfo
  {pages} {115411} (\bibinfo {year} {2015})}\BibitemShut {NoStop}%
\bibitem [{\citenamefont {Aydinol}\ and\ \citenamefont
  {Ceder}(1997)}]{aydinol1997a}%
  \BibitemOpen
  \bibfield  {author} {\bibinfo {author} {\bibfnamefont {M.~K.}\ \bibnamefont
  {Aydinol}}\ and\ \bibinfo {author} {\bibfnamefont {G.}~\bibnamefont
  {Ceder}},\ }\href {\doibase 10.1149/1.1838099} {\bibfield  {journal}
  {\bibinfo  {journal} {J. Electrochem. Soc.}\ }\textbf {\bibinfo {volume}
  {144}},\ \bibinfo {pages} {3832} (\bibinfo {year} {1997})}\BibitemShut
  {NoStop}%
\bibitem [{\citenamefont {Berg}\ \emph {et~al.}(1999)\citenamefont {Berg},
  \citenamefont {G{\"{o}}ransson}, \citenamefont {Nol{\"{a}}ng},\ and\
  \citenamefont {Thomas}}]{berg1999}%
  \BibitemOpen
  \bibfield  {author} {\bibinfo {author} {\bibfnamefont {H.}~\bibnamefont
  {Berg}}, \bibinfo {author} {\bibfnamefont {K.}~\bibnamefont
  {G{\"{o}}ransson}}, \bibinfo {author} {\bibfnamefont {B.}~\bibnamefont
  {Nol{\"{a}}ng}}, \ and\ \bibinfo {author} {\bibfnamefont {J.~O.}\
  \bibnamefont {Thomas}},\ }\href {\doibase 10.1039/a905575d} {\bibfield
  {journal} {\bibinfo  {journal} {J. Mater. Chem.}\ }\textbf {\bibinfo {volume}
  {9}},\ \bibinfo {pages} {2813} (\bibinfo {year} {1999})}\BibitemShut
  {NoStop}%
\bibitem [{\citenamefont {{Van der Ven}}\ \emph {et~al.}(2000)\citenamefont
  {{Van der Ven}}, \citenamefont {Marianetti}, \citenamefont {Morgan},\ and\
  \citenamefont {Ceder}}]{vanderven2000}%
  \BibitemOpen
  \bibfield  {author} {\bibinfo {author} {\bibfnamefont {A.}~\bibnamefont {{Van
  der Ven}}}, \bibinfo {author} {\bibfnamefont {C.}~\bibnamefont {Marianetti}},
  \bibinfo {author} {\bibfnamefont {D.}~\bibnamefont {Morgan}}, \ and\ \bibinfo
  {author} {\bibfnamefont {G.}~\bibnamefont {Ceder}},\ }\href {\doibase
  10.1016/S0167-2738(00)00326-X} {\bibfield  {journal} {\bibinfo  {journal}
  {Solid State Ionics}\ }\textbf {\bibinfo {volume} {135}},\ \bibinfo {pages}
  {21} (\bibinfo {year} {2000})}\BibitemShut {NoStop}%
\bibitem [{\citenamefont {Koyama}\ \emph {et~al.}(2003)\citenamefont {Koyama},
  \citenamefont {Tanaka}, \citenamefont {Adachi}, \citenamefont {Uchimoto},\
  and\ \citenamefont {Wakihara}}]{koyama2003}%
  \BibitemOpen
  \bibfield  {author} {\bibinfo {author} {\bibfnamefont {Y.}~\bibnamefont
  {Koyama}}, \bibinfo {author} {\bibfnamefont {I.}~\bibnamefont {Tanaka}},
  \bibinfo {author} {\bibfnamefont {H.}~\bibnamefont {Adachi}}, \bibinfo
  {author} {\bibfnamefont {Y.}~\bibnamefont {Uchimoto}}, \ and\ \bibinfo
  {author} {\bibfnamefont {M.}~\bibnamefont {Wakihara}},\ }\href {\doibase
  10.1149/1.1522720} {\bibfield  {journal} {\bibinfo  {journal} {J.
  Electrochem. Soc.}\ }\textbf {\bibinfo {volume} {150}},\ \bibinfo {pages}
  {A63} (\bibinfo {year} {2003})}\BibitemShut {NoStop}%
\bibitem [{\citenamefont {Shi}\ \emph {et~al.}(2003)\citenamefont {Shi},
  \citenamefont {Wang}, \citenamefont {Meng}, \citenamefont {Chen},\ and\
  \citenamefont {Huang}}]{shi2003}%
  \BibitemOpen
  \bibfield  {author} {\bibinfo {author} {\bibfnamefont {S.}~\bibnamefont
  {Shi}}, \bibinfo {author} {\bibfnamefont {D.-s.}\ \bibnamefont {Wang}},
  \bibinfo {author} {\bibfnamefont {S.}~\bibnamefont {Meng}}, \bibinfo {author}
  {\bibfnamefont {L.}~\bibnamefont {Chen}}, \ and\ \bibinfo {author}
  {\bibfnamefont {X.}~\bibnamefont {Huang}},\ }\href {\doibase
  10.1103/PhysRevB.67.115130} {\bibfield  {journal} {\bibinfo  {journal} {Phys.
  Rev. B}\ }\textbf {\bibinfo {volume} {67}},\ \bibinfo {pages} {115130}
  (\bibinfo {year} {2003})}\BibitemShut {NoStop}%
\bibitem [{\citenamefont {Benedek}\ and\ \citenamefont
  {Thackeray}(2011)}]{benedek2011}%
  \BibitemOpen
  \bibfield  {author} {\bibinfo {author} {\bibfnamefont {R.}~\bibnamefont
  {Benedek}}\ and\ \bibinfo {author} {\bibfnamefont {M.~M.}\ \bibnamefont
  {Thackeray}},\ }\href {\doibase 10.1103/PhysRevB.83.195439} {\bibfield
  {journal} {\bibinfo  {journal} {Phys. Rev. B}\ }\textbf {\bibinfo {volume}
  {83}},\ \bibinfo {pages} {195439} (\bibinfo {year} {2011})}\BibitemShut
  {NoStop}%
\bibitem [{\citenamefont {Tang}\ \emph {et~al.}(2014)\citenamefont {Tang},
  \citenamefont {Sun}, \citenamefont {Yang}, \citenamefont {Ben}, \citenamefont
  {Gu},\ and\ \citenamefont {Huang}}]{tang2014}%
  \BibitemOpen
  \bibfield  {author} {\bibinfo {author} {\bibfnamefont {D.}~\bibnamefont
  {Tang}}, \bibinfo {author} {\bibfnamefont {Y.}~\bibnamefont {Sun}}, \bibinfo
  {author} {\bibfnamefont {Z.}~\bibnamefont {Yang}}, \bibinfo {author}
  {\bibfnamefont {L.}~\bibnamefont {Ben}}, \bibinfo {author} {\bibfnamefont
  {L.}~\bibnamefont {Gu}}, \ and\ \bibinfo {author} {\bibfnamefont
  {X.}~\bibnamefont {Huang}},\ }\href {\doibase 10.1021/cm501125e} {\bibfield
  {journal} {\bibinfo  {journal} {Chem. Mater.}\ }\textbf {\bibinfo {volume}
  {26}},\ \bibinfo {pages} {3535} (\bibinfo {year} {2014})}\BibitemShut
  {NoStop}%
\bibitem [{\citenamefont {Ouyang}\ \emph {et~al.}(2010)\citenamefont {Ouyang},
  \citenamefont {Zeng}, \citenamefont {{\v{S}}ljivancanin},\ and\ \citenamefont
  {Baldereschi}}]{ouyang2010}%
  \BibitemOpen
  \bibfield  {author} {\bibinfo {author} {\bibfnamefont {C.~Y.}\ \bibnamefont
  {Ouyang}}, \bibinfo {author} {\bibfnamefont {X.~M.}\ \bibnamefont {Zeng}},
  \bibinfo {author} {\bibfnamefont {{\v{Z}}.}~\bibnamefont
  {{\v{S}}ljivancanin}}, \ and\ \bibinfo {author} {\bibfnamefont
  {A.}~\bibnamefont {Baldereschi}},\ }\href {\doibase 10.1021/jp911746g}
  {\bibfield  {journal} {\bibinfo  {journal} {J. Phys. Chem. C}\ }\textbf
  {\bibinfo {volume} {114}},\ \bibinfo {pages} {4756} (\bibinfo {year}
  {2010})}\BibitemShut {NoStop}%
\bibitem [{\citenamefont {Jaber-Ansari}\ \emph {et~al.}(2015)\citenamefont
  {Jaber-Ansari}, \citenamefont {Puntambekar}, \citenamefont {Kim},
  \citenamefont {Aykol}, \citenamefont {Luo}, \citenamefont {Wu}, \citenamefont
  {Myers}, \citenamefont {Iddir}, \citenamefont {Russell}, \citenamefont
  {Salda{\~{n}}a}, \citenamefont {Kumar}, \citenamefont {Thackeray},
  \citenamefont {Curtiss}, \citenamefont {Dravid}, \citenamefont {Wolverton},\
  and\ \citenamefont {Hersam}}]{jaber-ansari2015}%
  \BibitemOpen
  \bibfield  {author} {\bibinfo {author} {\bibfnamefont {L.}~\bibnamefont
  {Jaber-Ansari}}, \bibinfo {author} {\bibfnamefont {K.~P.}\ \bibnamefont
  {Puntambekar}}, \bibinfo {author} {\bibfnamefont {S.}~\bibnamefont {Kim}},
  \bibinfo {author} {\bibfnamefont {M.}~\bibnamefont {Aykol}}, \bibinfo
  {author} {\bibfnamefont {L.}~\bibnamefont {Luo}}, \bibinfo {author}
  {\bibfnamefont {J.}~\bibnamefont {Wu}}, \bibinfo {author} {\bibfnamefont
  {B.~D.}\ \bibnamefont {Myers}}, \bibinfo {author} {\bibfnamefont
  {H.}~\bibnamefont {Iddir}}, \bibinfo {author} {\bibfnamefont {J.~T.}\
  \bibnamefont {Russell}}, \bibinfo {author} {\bibfnamefont {S.~J.}\
  \bibnamefont {Salda{\~{n}}a}}, \bibinfo {author} {\bibfnamefont
  {R.}~\bibnamefont {Kumar}}, \bibinfo {author} {\bibfnamefont {M.~M.}\
  \bibnamefont {Thackeray}}, \bibinfo {author} {\bibfnamefont {L.~A.}\
  \bibnamefont {Curtiss}}, \bibinfo {author} {\bibfnamefont {V.~P.}\
  \bibnamefont {Dravid}}, \bibinfo {author} {\bibfnamefont {C.}~\bibnamefont
  {Wolverton}}, \ and\ \bibinfo {author} {\bibfnamefont {M.~C.}\ \bibnamefont
  {Hersam}},\ }\href {\doibase 10.1002/aenm.201500646} {\bibfield  {journal}
  {\bibinfo  {journal} {Adv. Energy Mater.}\ }\textbf {\bibinfo {volume} {5}},\
  \bibinfo {pages} {1} (\bibinfo {year} {2015})}\BibitemShut {NoStop}%
\bibitem [{\citenamefont {Kumar}\ \emph {et~al.}(2014)\citenamefont {Kumar},
  \citenamefont {Leung},\ and\ \citenamefont {Siegel}}]{kumar2014}%
  \BibitemOpen
  \bibfield  {author} {\bibinfo {author} {\bibfnamefont {N.}~\bibnamefont
  {Kumar}}, \bibinfo {author} {\bibfnamefont {K.}~\bibnamefont {Leung}}, \ and\
  \bibinfo {author} {\bibfnamefont {D.~J.}\ \bibnamefont {Siegel}},\ }\href
  {\doibase 10.1149/2.009408jes} {\bibfield  {journal} {\bibinfo  {journal} {J.
  Electrochem. Soc.}\ }\textbf {\bibinfo {volume} {161}},\ \bibinfo {pages}
  {E3059} (\bibinfo {year} {2014})}\BibitemShut {NoStop}%
\bibitem [{\citenamefont {Nakayama}\ \emph {et~al.}(2014)\citenamefont
  {Nakayama}, \citenamefont {Taki}, \citenamefont {Nakamura}, \citenamefont
  {Tokuda}, \citenamefont {Jalem},\ and\ \citenamefont
  {Kasuga}}]{nakayama2014}%
  \BibitemOpen
  \bibfield  {author} {\bibinfo {author} {\bibfnamefont {M.}~\bibnamefont
  {Nakayama}}, \bibinfo {author} {\bibfnamefont {H.}~\bibnamefont {Taki}},
  \bibinfo {author} {\bibfnamefont {T.}~\bibnamefont {Nakamura}}, \bibinfo
  {author} {\bibfnamefont {S.}~\bibnamefont {Tokuda}}, \bibinfo {author}
  {\bibfnamefont {R.}~\bibnamefont {Jalem}}, \ and\ \bibinfo {author}
  {\bibfnamefont {T.}~\bibnamefont {Kasuga}},\ }\href {\doibase
  10.1021/jp509232m} {\bibfield  {journal} {\bibinfo  {journal} {J. Phys. Chem.
  C}\ }\textbf {\bibinfo {volume} {118}},\ \bibinfo {pages} {27245} (\bibinfo
  {year} {2014})}\BibitemShut {NoStop}%
\bibitem [{\citenamefont {Chukalkin}\ \emph {et~al.}(2010)\citenamefont
  {Chukalkin}, \citenamefont {Teplykh}, \citenamefont {Pirogov},\ and\
  \citenamefont {Kellerman}}]{chukalkin2010}%
  \BibitemOpen
  \bibfield  {author} {\bibinfo {author} {\bibfnamefont {Y.~G.}\ \bibnamefont
  {Chukalkin}}, \bibinfo {author} {\bibfnamefont {A.~E.}\ \bibnamefont
  {Teplykh}}, \bibinfo {author} {\bibfnamefont {A.~N.}\ \bibnamefont
  {Pirogov}}, \ and\ \bibinfo {author} {\bibfnamefont {D.~G.}\ \bibnamefont
  {Kellerman}},\ }\href {\doibase 10.1134/S1063783410120164} {\bibfield
  {journal} {\bibinfo  {journal} {Phys. Solid State}\ }\textbf {\bibinfo
  {volume} {52}},\ \bibinfo {pages} {2545} (\bibinfo {year}
  {2010})}\BibitemShut {NoStop}%
\bibitem [{\citenamefont {Masquelier}\ \emph {et~al.}(1996)\citenamefont
  {Masquelier}, \citenamefont {Tabuchi}, \citenamefont {Ado}, \citenamefont
  {Kanno}, \citenamefont {Kobayashi}, \citenamefont {Maki}, \citenamefont
  {Nakamura},\ and\ \citenamefont {Goodenough}}]{masquelier1996}%
  \BibitemOpen
  \bibfield  {author} {\bibinfo {author} {\bibfnamefont {C.}~\bibnamefont
  {Masquelier}}, \bibinfo {author} {\bibfnamefont {M.}~\bibnamefont {Tabuchi}},
  \bibinfo {author} {\bibfnamefont {K.}~\bibnamefont {Ado}}, \bibinfo {author}
  {\bibfnamefont {R.}~\bibnamefont {Kanno}}, \bibinfo {author} {\bibfnamefont
  {Y.}~\bibnamefont {Kobayashi}}, \bibinfo {author} {\bibfnamefont
  {Y.}~\bibnamefont {Maki}}, \bibinfo {author} {\bibfnamefont {O.}~\bibnamefont
  {Nakamura}}, \ and\ \bibinfo {author} {\bibfnamefont {J.~B.}\ \bibnamefont
  {Goodenough}},\ }\href {\doibase 10.1006/jssc.1996.0176} {\bibfield
  {journal} {\bibinfo  {journal} {J. Solid State Chem.}\ }\textbf {\bibinfo
  {volume} {123}},\ \bibinfo {pages} {255} (\bibinfo {year}
  {1996})}\BibitemShut {NoStop}%
\bibitem [{\citenamefont {Goodenough}\ and\ \citenamefont
  {Loeb}(1955)}]{goodenough1955}%
  \BibitemOpen
  \bibfield  {author} {\bibinfo {author} {\bibfnamefont {J.~B.}\ \bibnamefont
  {Goodenough}}\ and\ \bibinfo {author} {\bibfnamefont {A.~L.}\ \bibnamefont
  {Loeb}},\ }\href {\doibase 10.1103/PhysRev.98.391} {\bibfield  {journal}
  {\bibinfo  {journal} {Phys. Rev.}\ }\textbf {\bibinfo {volume} {98}},\
  \bibinfo {pages} {391} (\bibinfo {year} {1955})}\BibitemShut {NoStop}%
\bibitem [{\citenamefont {Goodenough}(1955)}]{goodenough1955a}%
  \BibitemOpen
  \bibfield  {author} {\bibinfo {author} {\bibfnamefont {J.~B.}\ \bibnamefont
  {Goodenough}},\ }\href {\doibase 10.1103/PhysRev.100.564} {\bibfield
  {journal} {\bibinfo  {journal} {Phys. Rev.}\ }\textbf {\bibinfo {volume}
  {100}},\ \bibinfo {pages} {564} (\bibinfo {year} {1955})}\BibitemShut
  {NoStop}%
\bibitem [{\citenamefont {Goodenough}(1963)}]{goodenough1963}%
  \BibitemOpen
  \bibfield  {author} {\bibinfo {author} {\bibfnamefont {J.~B.}\ \bibnamefont
  {Goodenough}},\ }\href@noop {} {\emph {\bibinfo {title} {{Magnetism and the
  chemical bond - Interscience Monographs on Chemistry - Inorganic Chemistry
  Section - Volume 1}}}},\ edited by\ \bibinfo {editor} {\bibfnamefont {F.~A.}\
  \bibnamefont {Cotton}}\ (\bibinfo  {publisher} {Interscience Publishers a
  division of John Wiley {\&} Sons},\ \bibinfo {address} {New York-London},\
  \bibinfo {year} {1963})\ p.\ \bibinfo {pages} {393}\BibitemShut {NoStop}%
\bibitem [{\citenamefont {Kresse}\ and\ \citenamefont
  {Hafner}(1993)}]{kresse1993}%
  \BibitemOpen
  \bibfield  {author} {\bibinfo {author} {\bibfnamefont {G.}~\bibnamefont
  {Kresse}}\ and\ \bibinfo {author} {\bibfnamefont {J.}~\bibnamefont
  {Hafner}},\ }\href {\doibase 10.1103/PhysRevB.47.558} {\bibfield  {journal}
  {\bibinfo  {journal} {Phys. Rev. B}\ }\textbf {\bibinfo {volume} {47}},\
  \bibinfo {pages} {558} (\bibinfo {year} {1993})}\BibitemShut {NoStop}%
\bibitem [{\citenamefont {Kresse}\ and\ \citenamefont
  {Hafner}(1994)}]{kresse1994}%
  \BibitemOpen
  \bibfield  {author} {\bibinfo {author} {\bibfnamefont {G.}~\bibnamefont
  {Kresse}}\ and\ \bibinfo {author} {\bibfnamefont {J.}~\bibnamefont
  {Hafner}},\ }\href {\doibase 10.1103/PhysRevB.49.14251} {\bibfield  {journal}
  {\bibinfo  {journal} {Phys. Rev. B}\ }\textbf {\bibinfo {volume} {49}},\
  \bibinfo {pages} {14251} (\bibinfo {year} {1994})}\BibitemShut {NoStop}%
\bibitem [{\citenamefont {Kresse}\ and\ \citenamefont
  {Furthm{\"{u}}ller}(1996{\natexlab{a}})}]{kresse1996}%
  \BibitemOpen
  \bibfield  {author} {\bibinfo {author} {\bibfnamefont {G.}~\bibnamefont
  {Kresse}}\ and\ \bibinfo {author} {\bibfnamefont {J.}~\bibnamefont
  {Furthm{\"{u}}ller}},\ }\href {\doibase 10.1103/PhysRevB.54.11169} {\bibfield
   {journal} {\bibinfo  {journal} {Phys. Rev. B}\ }\textbf {\bibinfo {volume}
  {54}},\ \bibinfo {pages} {11169} (\bibinfo {year}
  {1996}{\natexlab{a}})}\BibitemShut {NoStop}%
\bibitem [{\citenamefont {Kresse}\ and\ \citenamefont
  {Furthm{\"{u}}ller}(1996{\natexlab{b}})}]{kresse1996a}%
  \BibitemOpen
  \bibfield  {author} {\bibinfo {author} {\bibfnamefont {G.}~\bibnamefont
  {Kresse}}\ and\ \bibinfo {author} {\bibfnamefont {J.}~\bibnamefont
  {Furthm{\"{u}}ller}},\ }\href {\doibase 10.1016/0927-0256(96)00008-0}
  {\bibfield  {journal} {\bibinfo  {journal} {Comput. Mater. Sci.}\ }\textbf
  {\bibinfo {volume} {6}},\ \bibinfo {pages} {15} (\bibinfo {year}
  {1996}{\natexlab{b}})}\BibitemShut {NoStop}%
\bibitem [{\citenamefont {Perdew}\ \emph {et~al.}(1996)\citenamefont {Perdew},
  \citenamefont {Burke},\ and\ \citenamefont {Ernzerhof}}]{Perdew1996}%
  \BibitemOpen
  \bibfield  {author} {\bibinfo {author} {\bibfnamefont {J.~P.}\ \bibnamefont
  {Perdew}}, \bibinfo {author} {\bibfnamefont {K.}~\bibnamefont {Burke}}, \
  and\ \bibinfo {author} {\bibfnamefont {M.}~\bibnamefont {Ernzerhof}},\ }\href
  {\doibase 10.1103/PhysRevLett.77.3865} {\bibfield  {journal} {\bibinfo
  {journal} {Phys. Rev. Lett.}\ }\textbf {\bibinfo {volume} {77}},\ \bibinfo
  {pages} {3865} (\bibinfo {year} {1996})}\BibitemShut {NoStop}%
\bibitem [{\citenamefont {Perdew}\ \emph {et~al.}(1997)\citenamefont {Perdew},
  \citenamefont {Burke},\ and\ \citenamefont {Ernzerhof}}]{Perdew1997}%
  \BibitemOpen
  \bibfield  {author} {\bibinfo {author} {\bibfnamefont {J.~P.}\ \bibnamefont
  {Perdew}}, \bibinfo {author} {\bibfnamefont {K.}~\bibnamefont {Burke}}, \
  and\ \bibinfo {author} {\bibfnamefont {M.}~\bibnamefont {Ernzerhof}},\ }\href
  {\doibase 10.1103/PhysRevLett.78.1396} {\bibfield  {journal} {\bibinfo
  {journal} {Phys. Rev. Lett.}\ }\textbf {\bibinfo {volume} {78}},\ \bibinfo
  {pages} {1396} (\bibinfo {year} {1997})}\BibitemShut {NoStop}%
\bibitem [{\citenamefont {Bl{\"{o}}chl}(1994)}]{blochl1994}%
  \BibitemOpen
  \bibfield  {author} {\bibinfo {author} {\bibfnamefont {P.~E.}\ \bibnamefont
  {Bl{\"{o}}chl}},\ }\href {\doibase 10.1103/PhysRevB.50.17953} {\bibfield
  {journal} {\bibinfo  {journal} {Phys. Rev. B}\ }\textbf {\bibinfo {volume}
  {50}},\ \bibinfo {pages} {17953} (\bibinfo {year} {1994})}\BibitemShut
  {NoStop}%
\bibitem [{\citenamefont {Kresse}\ and\ \citenamefont
  {Joubert}(1999)}]{kresse1999}%
  \BibitemOpen
  \bibfield  {author} {\bibinfo {author} {\bibfnamefont {G.}~\bibnamefont
  {Kresse}}\ and\ \bibinfo {author} {\bibfnamefont {D.}~\bibnamefont
  {Joubert}},\ }\href {\doibase 10.1103/PhysRevB.59.1758} {\bibfield  {journal}
  {\bibinfo  {journal} {Phys. Rev. B}\ }\textbf {\bibinfo {volume} {59}},\
  \bibinfo {pages} {1758} (\bibinfo {year} {1999})}\BibitemShut {NoStop}%
\bibitem [{\citenamefont {Monkhorst}\ and\ \citenamefont
  {Pack}(1976)}]{monkhorst1976}%
  \BibitemOpen
  \bibfield  {author} {\bibinfo {author} {\bibfnamefont {H.~J.}\ \bibnamefont
  {Monkhorst}}\ and\ \bibinfo {author} {\bibfnamefont {J.~D.}\ \bibnamefont
  {Pack}},\ }\href {\doibase 10.1103/PhysRevB.13.5188} {\bibfield  {journal}
  {\bibinfo  {journal} {Phys. Rev. B}\ }\textbf {\bibinfo {volume} {13}},\
  \bibinfo {pages} {5188} (\bibinfo {year} {1976})}\BibitemShut {NoStop}%
\bibitem [{\citenamefont {Mermin}(1965)}]{mermin1965}%
  \BibitemOpen
  \bibfield  {author} {\bibinfo {author} {\bibfnamefont {N.~D.}\ \bibnamefont
  {Mermin}},\ }\href {\doibase 10.1103/PhysRev.137.A1441} {\bibfield  {journal}
  {\bibinfo  {journal} {Phys. Rev.}\ }\textbf {\bibinfo {volume} {137}},\
  \bibinfo {pages} {A1441} (\bibinfo {year} {1965})}\BibitemShut {NoStop}%
\bibitem [{\citenamefont {Bl{\"{o}}chl}\ \emph {et~al.}(1994)\citenamefont
  {Bl{\"{o}}chl}, \citenamefont {Jepsen},\ and\ \citenamefont
  {Andersen}}]{blochl1994a}%
  \BibitemOpen
  \bibfield  {author} {\bibinfo {author} {\bibfnamefont {P.~E.}\ \bibnamefont
  {Bl{\"{o}}chl}}, \bibinfo {author} {\bibfnamefont {O.}~\bibnamefont
  {Jepsen}}, \ and\ \bibinfo {author} {\bibfnamefont {O.~K.}\ \bibnamefont
  {Andersen}},\ }\href {\doibase 10.1103/PhysRevB.49.16223} {\bibfield
  {journal} {\bibinfo  {journal} {Phys. Rev. B}\ }\textbf {\bibinfo {volume}
  {49}},\ \bibinfo {pages} {16223} (\bibinfo {year} {1994})}\BibitemShut
  {NoStop}%
\bibitem [{\citenamefont {Grimme}\ \emph {et~al.}(2010)\citenamefont {Grimme},
  \citenamefont {Antony}, \citenamefont {Ehrlich},\ and\ \citenamefont
  {Krieg}}]{grimme2010a}%
  \BibitemOpen
  \bibfield  {author} {\bibinfo {author} {\bibfnamefont {S.}~\bibnamefont
  {Grimme}}, \bibinfo {author} {\bibfnamefont {J.}~\bibnamefont {Antony}},
  \bibinfo {author} {\bibfnamefont {S.}~\bibnamefont {Ehrlich}}, \ and\
  \bibinfo {author} {\bibfnamefont {H.}~\bibnamefont {Krieg}},\ }\href
  {\doibase 10.1063/1.3382344} {\bibfield  {journal} {\bibinfo  {journal} {J.
  Chem. Phys.}\ }\textbf {\bibinfo {volume} {132}},\ \bibinfo {pages} {154104}
  (\bibinfo {year} {2010})}\BibitemShut {NoStop}%
\bibitem [{\citenamefont {Grimme}\ \emph {et~al.}(2011)\citenamefont {Grimme},
  \citenamefont {Ehrlich},\ and\ \citenamefont {Goerigk}}]{grimme2011}%
  \BibitemOpen
  \bibfield  {author} {\bibinfo {author} {\bibfnamefont {S.}~\bibnamefont
  {Grimme}}, \bibinfo {author} {\bibfnamefont {S.}~\bibnamefont {Ehrlich}}, \
  and\ \bibinfo {author} {\bibfnamefont {L.}~\bibnamefont {Goerigk}},\ }\href
  {\doibase 10.1002/jcc.21759} {\bibfield  {journal} {\bibinfo  {journal} {J.
  Comput. Chem.}\ }\textbf {\bibinfo {volume} {32}},\ \bibinfo {pages} {1456}
  (\bibinfo {year} {2011})}\BibitemShut {NoStop}%
\bibitem [{\citenamefont {Santos-Carballal}\ \emph {et~al.}(2014)\citenamefont
  {Santos-Carballal}, \citenamefont {Roldan}, \citenamefont {Grau-Crespo},\
  and\ \citenamefont {de~Leeuw}}]{santos-carballal2014}%
  \BibitemOpen
  \bibfield  {author} {\bibinfo {author} {\bibfnamefont {D.}~\bibnamefont
  {Santos-Carballal}}, \bibinfo {author} {\bibfnamefont {A.}~\bibnamefont
  {Roldan}}, \bibinfo {author} {\bibfnamefont {R.}~\bibnamefont {Grau-Crespo}},
  \ and\ \bibinfo {author} {\bibfnamefont {N.~H.}\ \bibnamefont {de~Leeuw}},\
  }\href {\doibase 10.1039/c4cp00529e} {\bibfield  {journal} {\bibinfo
  {journal} {Phys. Chem. Chem. Phys.}\ }\textbf {\bibinfo {volume} {16}},\
  \bibinfo {pages} {21082} (\bibinfo {year} {2014})}\BibitemShut {NoStop}%
\bibitem [{\citenamefont {Shields}\ \emph {et~al.}(2016)\citenamefont
  {Shields}, \citenamefont {Santos-Carballal},\ and\ \citenamefont
  {de~Leeuw}}]{shields2015}%
  \BibitemOpen
  \bibfield  {author} {\bibinfo {author} {\bibfnamefont {A.~E.}\ \bibnamefont
  {Shields}}, \bibinfo {author} {\bibfnamefont {D.}~\bibnamefont
  {Santos-Carballal}}, \ and\ \bibinfo {author} {\bibfnamefont {N.~H.}\
  \bibnamefont {de~Leeuw}},\ }\href {\doibase 10.1016/j.jnucmat.2016.02.009}
  {\bibfield  {journal} {\bibinfo  {journal} {J. Nucl. Mater.}\ }\textbf
  {\bibinfo {volume} {473}},\ \bibinfo {pages} {99} (\bibinfo {year}
  {2016})}\BibitemShut {NoStop}%
\bibitem [{\citenamefont {Santos-Carballal}\ \emph {et~al.}(2016)\citenamefont
  {Santos-Carballal}, \citenamefont {Roldan},\ and\ \citenamefont
  {de~Leeuw}}]{santos-carballal2016}%
  \BibitemOpen
  \bibfield  {author} {\bibinfo {author} {\bibfnamefont {D.}~\bibnamefont
  {Santos-Carballal}}, \bibinfo {author} {\bibfnamefont {A.}~\bibnamefont
  {Roldan}}, \ and\ \bibinfo {author} {\bibfnamefont {N.~H.}\ \bibnamefont
  {de~Leeuw}},\ }\href {\doibase 10.1021/acs.jpcc.6b00216} {\bibfield
  {journal} {\bibinfo  {journal} {J. Phys. Chem. C}\ }\textbf {\bibinfo
  {volume} {120}},\ \bibinfo {pages} {8616} (\bibinfo {year}
  {2016})}\BibitemShut {NoStop}%
\bibitem [{\citenamefont {Dzade}\ \emph {et~al.}(2017)\citenamefont {Dzade},
  \citenamefont {Roldan},\ and\ \citenamefont {de~Leeuw}}]{dzade2017}%
  \BibitemOpen
  \bibfield  {author} {\bibinfo {author} {\bibfnamefont {N.~Y.}\ \bibnamefont
  {Dzade}}, \bibinfo {author} {\bibfnamefont {A.}~\bibnamefont {Roldan}}, \
  and\ \bibinfo {author} {\bibfnamefont {N.~H.}\ \bibnamefont {de~Leeuw}},\
  }\href {\doibase 10.1021/acs.est.7b00107} {\bibfield  {journal} {\bibinfo
  {journal} {Environ. Sci. Technol.}\ }\textbf {\bibinfo {volume} {51}},\
  \bibinfo {pages} {3461} (\bibinfo {year} {2017})}\BibitemShut {NoStop}%
\bibitem [{\citenamefont {Santos-Carballal}\ \emph {et~al.}(2018)\citenamefont
  {Santos-Carballal}, \citenamefont {Roldan}, \citenamefont {Dzade},\ and\
  \citenamefont {de~Leeuw}}]{santos-carballal2017a}%
  \BibitemOpen
  \bibfield  {author} {\bibinfo {author} {\bibfnamefont {D.}~\bibnamefont
  {Santos-Carballal}}, \bibinfo {author} {\bibfnamefont {A.}~\bibnamefont
  {Roldan}}, \bibinfo {author} {\bibfnamefont {N.~Y.}\ \bibnamefont {Dzade}}, \
  and\ \bibinfo {author} {\bibfnamefont {N.~H.}\ \bibnamefont {de~Leeuw}},\
  }\href {\doibase 10.1098/rsta.2017.0065} {\bibfield  {journal} {\bibinfo
  {journal} {Philos. Trans. R. Soc. A Math. Phys. Eng. Sci.}\ }\textbf
  {\bibinfo {volume} {376}},\ \bibinfo {pages} {20170065} (\bibinfo {year}
  {2018})}\BibitemShut {NoStop}%
\bibitem [{\citenamefont {Anisimov}\ \emph {et~al.}(1992)\citenamefont
  {Anisimov}, \citenamefont {Korotin}, \citenamefont {Zaanen},\ and\
  \citenamefont {Andersen}}]{anisimov1992}%
  \BibitemOpen
  \bibfield  {author} {\bibinfo {author} {\bibfnamefont {V.~I.}\ \bibnamefont
  {Anisimov}}, \bibinfo {author} {\bibfnamefont {M.~A.}\ \bibnamefont
  {Korotin}}, \bibinfo {author} {\bibfnamefont {J.}~\bibnamefont {Zaanen}}, \
  and\ \bibinfo {author} {\bibfnamefont {O.~K.}\ \bibnamefont {Andersen}},\
  }\href {\doibase 10.1103/PhysRevLett.68.345} {\bibfield  {journal} {\bibinfo
  {journal} {Phys. Rev. Lett.}\ }\textbf {\bibinfo {volume} {68}},\ \bibinfo
  {pages} {345} (\bibinfo {year} {1992})}\BibitemShut {NoStop}%
\bibitem [{\citenamefont {Dudarev}\ \emph {et~al.}(1998)\citenamefont
  {Dudarev}, \citenamefont {Botton}, \citenamefont {Savrasov}, \citenamefont
  {Humphreys},\ and\ \citenamefont {Sutton}}]{dudarev1998}%
  \BibitemOpen
  \bibfield  {author} {\bibinfo {author} {\bibfnamefont {S.~L.}\ \bibnamefont
  {Dudarev}}, \bibinfo {author} {\bibfnamefont {G.~A.}\ \bibnamefont {Botton}},
  \bibinfo {author} {\bibfnamefont {S.~Y.}\ \bibnamefont {Savrasov}}, \bibinfo
  {author} {\bibfnamefont {C.~J.}\ \bibnamefont {Humphreys}}, \ and\ \bibinfo
  {author} {\bibfnamefont {A.~P.}\ \bibnamefont {Sutton}},\ }\href {\doibase
  10.1103/PhysRevB.57.1505} {\bibfield  {journal} {\bibinfo  {journal} {Phys.
  Rev. B}\ }\textbf {\bibinfo {volume} {57}},\ \bibinfo {pages} {1505}
  (\bibinfo {year} {1998})}\BibitemShut {NoStop}%
\bibitem [{\citenamefont {Heyd}\ \emph {et~al.}(2003)\citenamefont {Heyd},
  \citenamefont {Scuseria},\ and\ \citenamefont {Ernzerhof}}]{heyd2003}%
  \BibitemOpen
  \bibfield  {author} {\bibinfo {author} {\bibfnamefont {J.}~\bibnamefont
  {Heyd}}, \bibinfo {author} {\bibfnamefont {G.~E.}\ \bibnamefont {Scuseria}},
  \ and\ \bibinfo {author} {\bibfnamefont {M.}~\bibnamefont {Ernzerhof}},\
  }\href {\doibase 10.1063/1.1564060} {\bibfield  {journal} {\bibinfo
  {journal} {J. Chem. Phys.}\ }\textbf {\bibinfo {volume} {118}},\ \bibinfo
  {pages} {8207} (\bibinfo {year} {2003})}\BibitemShut {NoStop}%
\bibitem [{\citenamefont {Heyd}\ \emph {et~al.}(2006)\citenamefont {Heyd},
  \citenamefont {Scuseria},\ and\ \citenamefont {Ernzerhof}}]{heyd2006}%
  \BibitemOpen
  \bibfield  {author} {\bibinfo {author} {\bibfnamefont {J.}~\bibnamefont
  {Heyd}}, \bibinfo {author} {\bibfnamefont {G.~E.}\ \bibnamefont {Scuseria}},
  \ and\ \bibinfo {author} {\bibfnamefont {M.}~\bibnamefont {Ernzerhof}},\
  }\href {\doibase 10.1063/1.2204597} {\bibfield  {journal} {\bibinfo
  {journal} {J. Chem. Phys.}\ }\textbf {\bibinfo {volume} {124}},\ \bibinfo
  {pages} {219906} (\bibinfo {year} {2006})}\BibitemShut {NoStop}%
\bibitem [{\citenamefont {Heyd}\ and\ \citenamefont
  {Scuseria}(2004{\natexlab{a}})}]{heyd2004}%
  \BibitemOpen
  \bibfield  {author} {\bibinfo {author} {\bibfnamefont {J.}~\bibnamefont
  {Heyd}}\ and\ \bibinfo {author} {\bibfnamefont {G.~E.}\ \bibnamefont
  {Scuseria}},\ }\href {\doibase 10.1063/1.1668634} {\bibfield  {journal}
  {\bibinfo  {journal} {J. Chem. Phys.}\ }\textbf {\bibinfo {volume} {120}},\
  \bibinfo {pages} {7274} (\bibinfo {year} {2004}{\natexlab{a}})}\BibitemShut
  {NoStop}%
\bibitem [{\citenamefont {Heyd}\ and\ \citenamefont
  {Scuseria}(2004{\natexlab{b}})}]{heyd2004a}%
  \BibitemOpen
  \bibfield  {author} {\bibinfo {author} {\bibfnamefont {J.}~\bibnamefont
  {Heyd}}\ and\ \bibinfo {author} {\bibfnamefont {G.~E.}\ \bibnamefont
  {Scuseria}},\ }\href {\doibase 10.1063/1.1760074} {\bibfield  {journal}
  {\bibinfo  {journal} {J. Chem. Phys.}\ }\textbf {\bibinfo {volume} {121}},\
  \bibinfo {pages} {1187} (\bibinfo {year} {2004}{\natexlab{b}})}\BibitemShut
  {NoStop}%
\bibitem [{\citenamefont {Heyd}\ \emph {et~al.}(2005)\citenamefont {Heyd},
  \citenamefont {Peralta}, \citenamefont {Scuseria},\ and\ \citenamefont
  {Martin}}]{heyd2005}%
  \BibitemOpen
  \bibfield  {author} {\bibinfo {author} {\bibfnamefont {J.}~\bibnamefont
  {Heyd}}, \bibinfo {author} {\bibfnamefont {J.~E.}\ \bibnamefont {Peralta}},
  \bibinfo {author} {\bibfnamefont {G.~E.}\ \bibnamefont {Scuseria}}, \ and\
  \bibinfo {author} {\bibfnamefont {R.~L.}\ \bibnamefont {Martin}},\ }\href
  {\doibase 10.1063/1.2085170} {\bibfield  {journal} {\bibinfo  {journal} {J.
  Chem. Phys.}\ }\textbf {\bibinfo {volume} {123}},\ \bibinfo {pages} {174101}
  (\bibinfo {year} {2005})}\BibitemShut {NoStop}%
\bibitem [{\citenamefont {Peralta}\ \emph {et~al.}(2006)\citenamefont
  {Peralta}, \citenamefont {Heyd}, \citenamefont {Scuseria},\ and\
  \citenamefont {Martin}}]{peralta2006}%
  \BibitemOpen
  \bibfield  {author} {\bibinfo {author} {\bibfnamefont {J.~E.}\ \bibnamefont
  {Peralta}}, \bibinfo {author} {\bibfnamefont {J.}~\bibnamefont {Heyd}},
  \bibinfo {author} {\bibfnamefont {G.~E.}\ \bibnamefont {Scuseria}}, \ and\
  \bibinfo {author} {\bibfnamefont {R.~L.}\ \bibnamefont {Martin}},\ }\href
  {\doibase 10.1103/PhysRevB.74.073101} {\bibfield  {journal} {\bibinfo
  {journal} {Phys. Rev. B}\ }\textbf {\bibinfo {volume} {74}},\ \bibinfo
  {pages} {073101} (\bibinfo {year} {2006})}\BibitemShut {NoStop}%
\bibitem [{\citenamefont {Krukau}\ \emph {et~al.}(2006)\citenamefont {Krukau},
  \citenamefont {Vydrov}, \citenamefont {Izmaylov},\ and\ \citenamefont
  {Scuseria}}]{krukau2006}%
  \BibitemOpen
  \bibfield  {author} {\bibinfo {author} {\bibfnamefont {A.~V.}\ \bibnamefont
  {Krukau}}, \bibinfo {author} {\bibfnamefont {O.~A.}\ \bibnamefont {Vydrov}},
  \bibinfo {author} {\bibfnamefont {A.~F.}\ \bibnamefont {Izmaylov}}, \ and\
  \bibinfo {author} {\bibfnamefont {G.~E.}\ \bibnamefont {Scuseria}},\ }\href
  {\doibase 10.1063/1.2404663} {\bibfield  {journal} {\bibinfo  {journal} {J.
  Chem. Phys.}\ }\textbf {\bibinfo {volume} {125}},\ \bibinfo {pages} {224106}
  (\bibinfo {year} {2006})}\BibitemShut {NoStop}%
\bibitem [{\citenamefont {Henkelman}\ \emph {et~al.}(2006)\citenamefont
  {Henkelman}, \citenamefont {Arnaldsson},\ and\ \citenamefont
  {J{\'{o}}nsson}}]{henkelman2006}%
  \BibitemOpen
  \bibfield  {author} {\bibinfo {author} {\bibfnamefont {G.}~\bibnamefont
  {Henkelman}}, \bibinfo {author} {\bibfnamefont {A.}~\bibnamefont
  {Arnaldsson}}, \ and\ \bibinfo {author} {\bibfnamefont {H.}~\bibnamefont
  {J{\'{o}}nsson}},\ }\href {\doibase 10.1016/j.commatsci.2005.04.010}
  {\bibfield  {journal} {\bibinfo  {journal} {Comput. Mater. Sci.}\ }\textbf
  {\bibinfo {volume} {36}},\ \bibinfo {pages} {354} (\bibinfo {year}
  {2006})}\BibitemShut {NoStop}%
\bibitem [{\citenamefont {Sanville}\ \emph {et~al.}(2007)\citenamefont
  {Sanville}, \citenamefont {Kenny}, \citenamefont {Smith},\ and\ \citenamefont
  {Henkelman}}]{sanville2007}%
  \BibitemOpen
  \bibfield  {author} {\bibinfo {author} {\bibfnamefont {E.}~\bibnamefont
  {Sanville}}, \bibinfo {author} {\bibfnamefont {S.~D.}\ \bibnamefont {Kenny}},
  \bibinfo {author} {\bibfnamefont {R.}~\bibnamefont {Smith}}, \ and\ \bibinfo
  {author} {\bibfnamefont {G.}~\bibnamefont {Henkelman}},\ }\href {\doibase
  10.1002/jcc.20575} {\bibfield  {journal} {\bibinfo  {journal} {J. Comput.
  Chem.}\ }\textbf {\bibinfo {volume} {28}},\ \bibinfo {pages} {899} (\bibinfo
  {year} {2007})}\BibitemShut {NoStop}%
\bibitem [{\citenamefont {Tang}\ \emph {et~al.}(2009)\citenamefont {Tang},
  \citenamefont {Sanville},\ and\ \citenamefont {Henkelman}}]{tang2009}%
  \BibitemOpen
  \bibfield  {author} {\bibinfo {author} {\bibfnamefont {W.}~\bibnamefont
  {Tang}}, \bibinfo {author} {\bibfnamefont {E.}~\bibnamefont {Sanville}}, \
  and\ \bibinfo {author} {\bibfnamefont {G.}~\bibnamefont {Henkelman}},\ }\href
  {\doibase 10.1088/0953-8984/21/8/084204} {\bibfield  {journal} {\bibinfo
  {journal} {J. Phys. Condens. Matter}\ }\textbf {\bibinfo {volume} {21}},\
  \bibinfo {pages} {084204} (\bibinfo {year} {2009})}\BibitemShut {NoStop}%
\bibitem [{\citenamefont {Grau-Crespo}\ \emph {et~al.}(2007)\citenamefont
  {Grau-Crespo}, \citenamefont {Hamad}, \citenamefont {Catlow},\ and\
  \citenamefont {de~Leeuw}}]{grau-crespo2007a}%
  \BibitemOpen
  \bibfield  {author} {\bibinfo {author} {\bibfnamefont {R.}~\bibnamefont
  {Grau-Crespo}}, \bibinfo {author} {\bibfnamefont {S.}~\bibnamefont {Hamad}},
  \bibinfo {author} {\bibfnamefont {C.~R.~A.}\ \bibnamefont {Catlow}}, \ and\
  \bibinfo {author} {\bibfnamefont {N.~H.}\ \bibnamefont {de~Leeuw}},\ }\href
  {\doibase 10.1088/0953-8984/19/25/256201} {\bibfield  {journal} {\bibinfo
  {journal} {J. Phys. Condens. Matter}\ }\textbf {\bibinfo {volume} {19}},\
  \bibinfo {pages} {256201} (\bibinfo {year} {2007})}\BibitemShut {NoStop}%
\bibitem [{\citenamefont {Navrotsky}\ and\ \citenamefont
  {Kleppa}(1967)}]{navrotsky1967}%
  \BibitemOpen
  \bibfield  {author} {\bibinfo {author} {\bibfnamefont {A.}~\bibnamefont
  {Navrotsky}}\ and\ \bibinfo {author} {\bibfnamefont {O.~J.}\ \bibnamefont
  {Kleppa}},\ }\href {\doibase 10.1016/0022-1902(67)80008-3} {\bibfield
  {journal} {\bibinfo  {journal} {J. Inorg. Nucl. Chem.}\ }\textbf {\bibinfo
  {volume} {29}},\ \bibinfo {pages} {2701} (\bibinfo {year}
  {1967})}\BibitemShut {NoStop}%
\bibitem [{\citenamefont {Tielens}\ \emph {et~al.}(2006)\citenamefont
  {Tielens}, \citenamefont {Calatayud}, \citenamefont {Franco}, \citenamefont
  {Recio}, \citenamefont {P{\'{e}}rez-Ram{\'{i}}rez},\ and\ \citenamefont
  {Minot}}]{tielens2006}%
  \BibitemOpen
  \bibfield  {author} {\bibinfo {author} {\bibfnamefont {F.}~\bibnamefont
  {Tielens}}, \bibinfo {author} {\bibfnamefont {M.}~\bibnamefont {Calatayud}},
  \bibinfo {author} {\bibfnamefont {R.}~\bibnamefont {Franco}}, \bibinfo
  {author} {\bibfnamefont {J.~M.}\ \bibnamefont {Recio}}, \bibinfo {author}
  {\bibfnamefont {J.}~\bibnamefont {P{\'{e}}rez-Ram{\'{i}}rez}}, \ and\
  \bibinfo {author} {\bibfnamefont {C.}~\bibnamefont {Minot}},\ }\href
  {\doibase 10.1021/jp053375l} {\bibfield  {journal} {\bibinfo  {journal} {J.
  Phys. Chem. B}\ }\textbf {\bibinfo {volume} {110}},\ \bibinfo {pages} {988}
  (\bibinfo {year} {2006})}\BibitemShut {NoStop}%
\bibitem [{\citenamefont {Palin}\ \emph {et~al.}(2008)\citenamefont {Palin},
  \citenamefont {Walker},\ and\ \citenamefont {Harrison}}]{palin2008}%
  \BibitemOpen
  \bibfield  {author} {\bibinfo {author} {\bibfnamefont {E.~J.}\ \bibnamefont
  {Palin}}, \bibinfo {author} {\bibfnamefont {A.~M.}\ \bibnamefont {Walker}}, \
  and\ \bibinfo {author} {\bibfnamefont {R.~J.}\ \bibnamefont {Harrison}},\
  }\href {\doibase 10.2138/am.2008.2896} {\bibfield  {journal} {\bibinfo
  {journal} {Am. Mineral.}\ }\textbf {\bibinfo {volume} {93}},\ \bibinfo
  {pages} {1363} (\bibinfo {year} {2008})}\BibitemShut {NoStop}%
\bibitem [{\citenamefont {Seko}\ \emph {et~al.}(2010)\citenamefont {Seko},
  \citenamefont {Oba},\ and\ \citenamefont {Tanaka}}]{seko2010}%
  \BibitemOpen
  \bibfield  {author} {\bibinfo {author} {\bibfnamefont {A.}~\bibnamefont
  {Seko}}, \bibinfo {author} {\bibfnamefont {F.}~\bibnamefont {Oba}}, \ and\
  \bibinfo {author} {\bibfnamefont {I.}~\bibnamefont {Tanaka}},\ }\href
  {\doibase 10.1103/PhysRevB.81.054114} {\bibfield  {journal} {\bibinfo
  {journal} {Phys. Rev. B}\ }\textbf {\bibinfo {volume} {81}},\ \bibinfo
  {pages} {054114} (\bibinfo {year} {2010})}\BibitemShut {NoStop}%
\bibitem [{\citenamefont {Ndione}\ \emph {et~al.}(2014)\citenamefont {Ndione},
  \citenamefont {Shi}, \citenamefont {Stevanovic}, \citenamefont {Lany},
  \citenamefont {Zakutayev}, \citenamefont {Parilla}, \citenamefont {Perkins},
  \citenamefont {Berry}, \citenamefont {Ginley},\ and\ \citenamefont
  {Toney}}]{ndione2014}%
  \BibitemOpen
  \bibfield  {author} {\bibinfo {author} {\bibfnamefont {P.~F.}\ \bibnamefont
  {Ndione}}, \bibinfo {author} {\bibfnamefont {Y.}~\bibnamefont {Shi}},
  \bibinfo {author} {\bibfnamefont {V.}~\bibnamefont {Stevanovic}}, \bibinfo
  {author} {\bibfnamefont {S.}~\bibnamefont {Lany}}, \bibinfo {author}
  {\bibfnamefont {A.}~\bibnamefont {Zakutayev}}, \bibinfo {author}
  {\bibfnamefont {P.~A.}\ \bibnamefont {Parilla}}, \bibinfo {author}
  {\bibfnamefont {J.~D.}\ \bibnamefont {Perkins}}, \bibinfo {author}
  {\bibfnamefont {J.~J.}\ \bibnamefont {Berry}}, \bibinfo {author}
  {\bibfnamefont {D.~S.}\ \bibnamefont {Ginley}}, \ and\ \bibinfo {author}
  {\bibfnamefont {M.~F.}\ \bibnamefont {Toney}},\ }\href {\doibase
  10.1002/adfm.201302535} {\bibfield  {journal} {\bibinfo  {journal} {Adv.
  Funct. Mater.}\ }\textbf {\bibinfo {volume} {24}},\ \bibinfo {pages} {610}
  (\bibinfo {year} {2014})}\BibitemShut {NoStop}%
\bibitem [{\citenamefont {Seminovski}\ \emph {et~al.}(2012)\citenamefont
  {Seminovski}, \citenamefont {Palacios}, \citenamefont {Wahnón},\ and\
  \citenamefont {Grau-Crespo}}]{seminovski2012}%
  \BibitemOpen
  \bibfield  {author} {\bibinfo {author} {\bibfnamefont {Y.}~\bibnamefont
  {Seminovski}}, \bibinfo {author} {\bibfnamefont {P.}~\bibnamefont
  {Palacios}}, \bibinfo {author} {\bibfnamefont {P.}~\bibnamefont {Wahnón}},
  \ and\ \bibinfo {author} {\bibfnamefont {R.}~\bibnamefont {Grau-Crespo}},\
  }\href {\doibase 10.1063/1.3692780} {\bibfield  {journal} {\bibinfo
  {journal} {Appl. Phys. Lett.}\ }\textbf {\bibinfo {volume} {100}},\ \bibinfo
  {pages} {102112} (\bibinfo {year} {2012})}\BibitemShut {NoStop}%
\bibitem [{\citenamefont {Santos-Carballal}\ \emph {et~al.}(2015)\citenamefont
  {Santos-Carballal}, \citenamefont {Roldan}, \citenamefont {Grau-Crespo},\
  and\ \citenamefont {de~Leeuw}}]{santos-carballal2015b}%
  \BibitemOpen
  \bibfield  {author} {\bibinfo {author} {\bibfnamefont {D.}~\bibnamefont
  {Santos-Carballal}}, \bibinfo {author} {\bibfnamefont {A.}~\bibnamefont
  {Roldan}}, \bibinfo {author} {\bibfnamefont {R.}~\bibnamefont {Grau-Crespo}},
  \ and\ \bibinfo {author} {\bibfnamefont {N.~H.}\ \bibnamefont {de~Leeuw}},\
  }\href {\doibase 10.1103/PhysRevB.91.195106} {\bibfield  {journal} {\bibinfo
  {journal} {Phys. Rev. B}\ }\textbf {\bibinfo {volume} {91}},\ \bibinfo
  {pages} {195106} (\bibinfo {year} {2015})},\ \Eprint
  {http://arxiv.org/abs/1504.00268} {arXiv:1504.00268} \BibitemShut {NoStop}%
\bibitem [{\citenamefont {Wei}\ and\ \citenamefont {Zhang}(2001)}]{wei2001}%
  \BibitemOpen
  \bibfield  {author} {\bibinfo {author} {\bibfnamefont {S.-H.}\ \bibnamefont
  {Wei}}\ and\ \bibinfo {author} {\bibfnamefont {S.~B.}\ \bibnamefont
  {Zhang}},\ }\href {\doibase 10.1103/PhysRevB.63.045112} {\bibfield  {journal}
  {\bibinfo  {journal} {Phys. Rev. B}\ }\textbf {\bibinfo {volume} {63}},\
  \bibinfo {pages} {045112} (\bibinfo {year} {2001})}\BibitemShut {NoStop}%
\bibitem [{\citenamefont {Walsh}\ \emph {et~al.}(2007)\citenamefont {Walsh},
  \citenamefont {Wei}, \citenamefont {Yan}, \citenamefont {Al-Jassim},
  \citenamefont {Turner}, \citenamefont {Woodhouse},\ and\ \citenamefont
  {Parkinson}}]{walsh2007}%
  \BibitemOpen
  \bibfield  {author} {\bibinfo {author} {\bibfnamefont {A.}~\bibnamefont
  {Walsh}}, \bibinfo {author} {\bibfnamefont {S.-H.}\ \bibnamefont {Wei}},
  \bibinfo {author} {\bibfnamefont {Y.}~\bibnamefont {Yan}}, \bibinfo {author}
  {\bibfnamefont {M.~M.}\ \bibnamefont {Al-Jassim}}, \bibinfo {author}
  {\bibfnamefont {J.~A.}\ \bibnamefont {Turner}}, \bibinfo {author}
  {\bibfnamefont {M.}~\bibnamefont {Woodhouse}}, \ and\ \bibinfo {author}
  {\bibfnamefont {B.~A.}\ \bibnamefont {Parkinson}},\ }\href {\doibase
  10.1103/PhysRevB.76.165119} {\bibfield  {journal} {\bibinfo  {journal} {Phys.
  Rev. B}\ }\textbf {\bibinfo {volume} {76}},\ \bibinfo {pages} {165119}
  (\bibinfo {year} {2007})}\BibitemShut {NoStop}%
\bibitem [{\citenamefont {Fritsch}\ and\ \citenamefont
  {Ederer}(2010)}]{fritsch2010}%
  \BibitemOpen
  \bibfield  {author} {\bibinfo {author} {\bibfnamefont {D.}~\bibnamefont
  {Fritsch}}\ and\ \bibinfo {author} {\bibfnamefont {C.}~\bibnamefont
  {Ederer}},\ }\href {\doibase 10.1103/PhysRevB.82.104117} {\bibfield
  {journal} {\bibinfo  {journal} {Phys. Rev. B}\ }\textbf {\bibinfo {volume}
  {82}},\ \bibinfo {pages} {104117} (\bibinfo {year} {2010})},\ \Eprint
  {http://arxiv.org/abs/1006.5080} {arXiv:1006.5080} \BibitemShut {NoStop}%
\bibitem [{\citenamefont {Fritsch}\ and\ \citenamefont
  {Ederer}(2011)}]{fritsch2011b}%
  \BibitemOpen
  \bibfield  {author} {\bibinfo {author} {\bibfnamefont {D.}~\bibnamefont
  {Fritsch}}\ and\ \bibinfo {author} {\bibfnamefont {C.}~\bibnamefont
  {Ederer}},\ }\href {\doibase 10.1088/1742-6596/292/1/012014} {\bibfield
  {journal} {\bibinfo  {journal} {J. Phys. Conf. Ser.}\ }\textbf {\bibinfo
  {volume} {292}},\ \bibinfo {pages} {012014} (\bibinfo {year}
  {2011})}\BibitemShut {NoStop}%
\bibitem [{\citenamefont {Brabers}(1995)}]{brabers1995}%
  \BibitemOpen
  \bibfield  {author} {\bibinfo {author} {\bibfnamefont {V.~A.~M.}\
  \bibnamefont {Brabers}},\ }in\ \href@noop {} {\emph {\bibinfo {booktitle}
  {Handb. Magn. Mater. Vol. 8}}},\ \bibinfo {editor} {edited by\ \bibinfo
  {editor} {\bibfnamefont {K.}~\bibnamefont {Buschow}}}\ (\bibinfo  {publisher}
  {Elsevier Science B.V.},\ \bibinfo {address} {Amsterdam},\ \bibinfo {year}
  {1995})\ Chap.~\bibinfo {chapter} {3}, pp.\ \bibinfo {pages}
  {189--324}\BibitemShut {NoStop}%
\bibitem [{\citenamefont {Kriessman}\ and\ \citenamefont
  {Harrison}(1956)}]{kriessman1956}%
  \BibitemOpen
  \bibfield  {author} {\bibinfo {author} {\bibfnamefont {C.~J.}\ \bibnamefont
  {Kriessman}}\ and\ \bibinfo {author} {\bibfnamefont {S.~E.}\ \bibnamefont
  {Harrison}},\ }\href {\doibase 10.1103/PhysRev.103.857} {\bibfield  {journal}
  {\bibinfo  {journal} {Phys. Rev.}\ }\textbf {\bibinfo {volume} {103}},\
  \bibinfo {pages} {857} (\bibinfo {year} {1956})}\BibitemShut {NoStop}%
\bibitem [{\citenamefont {O'Neill}\ and\ \citenamefont
  {Navrotsky}(1983)}]{o'neill1983}%
  \BibitemOpen
  \bibfield  {author} {\bibinfo {author} {\bibfnamefont {H.~S.~C.}\
  \bibnamefont {O'Neill}}\ and\ \bibinfo {author} {\bibfnamefont
  {A.}~\bibnamefont {Navrotsky}},\ }\href@noop {} {\bibfield  {journal}
  {\bibinfo  {journal} {Am. Mineral.}\ }\textbf {\bibinfo {volume} {68}},\
  \bibinfo {pages} {181} (\bibinfo {year} {1983})}\BibitemShut {NoStop}%
\bibitem [{\citenamefont {Hill}\ \emph {et~al.}(1979)\citenamefont {Hill},
  \citenamefont {Craig},\ and\ \citenamefont {Gibbs}}]{hill1979}%
  \BibitemOpen
  \bibfield  {author} {\bibinfo {author} {\bibfnamefont {R.~J.}\ \bibnamefont
  {Hill}}, \bibinfo {author} {\bibfnamefont {J.~R.}\ \bibnamefont {Craig}}, \
  and\ \bibinfo {author} {\bibfnamefont {G.~V.}\ \bibnamefont {Gibbs}},\ }\href
  {\doibase 10.1007/BF00307535} {\bibfield  {journal} {\bibinfo  {journal}
  {Phys. Chem. Miner.}\ }\textbf {\bibinfo {volume} {4}},\ \bibinfo {pages}
  {317} (\bibinfo {year} {1979})}\BibitemShut {NoStop}%
\bibitem [{\citenamefont {Shannon}(1976)}]{shannon1976}%
  \BibitemOpen
  \bibfield  {author} {\bibinfo {author} {\bibfnamefont {R.~D.}\ \bibnamefont
  {Shannon}},\ }\href {\doibase 10.1107/S0567739476001551} {\bibfield
  {journal} {\bibinfo  {journal} {Acta Crystallogr. Sect. A}\ }\textbf
  {\bibinfo {volume} {32}},\ \bibinfo {pages} {751} (\bibinfo {year}
  {1976})}\BibitemShut {NoStop}%
\bibitem [{\citenamefont {McClure}(1957)}]{mcclure1957}%
  \BibitemOpen
  \bibfield  {author} {\bibinfo {author} {\bibfnamefont {D.~S.}\ \bibnamefont
  {McClure}},\ }\href {\doibase 10.1016/0022-3697(57)90034-3} {\bibfield
  {journal} {\bibinfo  {journal} {J. Phys. Chem. Solids}\ }\textbf {\bibinfo
  {volume} {3}},\ \bibinfo {pages} {311} (\bibinfo {year} {1957})}\BibitemShut
  {NoStop}%
\bibitem [{\citenamefont {Dunitz}\ and\ \citenamefont
  {Orgel}(1957)}]{dunitz1957}%
  \BibitemOpen
  \bibfield  {author} {\bibinfo {author} {\bibfnamefont {J.~D.}\ \bibnamefont
  {Dunitz}}\ and\ \bibinfo {author} {\bibfnamefont {L.~E.}\ \bibnamefont
  {Orgel}},\ }\href {\doibase 10.1016/0022-3697(57)90035-5} {\bibfield
  {journal} {\bibinfo  {journal} {J. Phys. Chem. Solids}\ }\textbf {\bibinfo
  {volume} {3}},\ \bibinfo {pages} {318} (\bibinfo {year} {1957})}\BibitemShut
  {NoStop}%
\bibitem [{\citenamefont {Roldan}\ \emph {et~al.}(2013)\citenamefont {Roldan},
  \citenamefont {Santos-Carballal},\ and\ \citenamefont
  {de~Leeuw}}]{roldan2013}%
  \BibitemOpen
  \bibfield  {author} {\bibinfo {author} {\bibfnamefont {A.}~\bibnamefont
  {Roldan}}, \bibinfo {author} {\bibfnamefont {D.}~\bibnamefont
  {Santos-Carballal}}, \ and\ \bibinfo {author} {\bibfnamefont {N.~H.}\
  \bibnamefont {de~Leeuw}},\ }\href {\doibase 10.1063/1.4807614} {\bibfield
  {journal} {\bibinfo  {journal} {J. Chem. Phys.}\ }\textbf {\bibinfo {volume}
  {138}},\ \bibinfo {pages} {204712} (\bibinfo {year} {2013})}\BibitemShut
  {NoStop}%
\bibitem [{\citenamefont {Zhou}\ \emph {et~al.}(2004)\citenamefont {Zhou},
  \citenamefont {Cococcioni}, \citenamefont {Marianetti}, \citenamefont
  {Morgan},\ and\ \citenamefont {Ceder}}]{zhou2004}%
  \BibitemOpen
  \bibfield  {author} {\bibinfo {author} {\bibfnamefont {F.}~\bibnamefont
  {Zhou}}, \bibinfo {author} {\bibfnamefont {M.}~\bibnamefont {Cococcioni}},
  \bibinfo {author} {\bibfnamefont {C.~A.}\ \bibnamefont {Marianetti}},
  \bibinfo {author} {\bibfnamefont {D.}~\bibnamefont {Morgan}}, \ and\ \bibinfo
  {author} {\bibfnamefont {G.}~\bibnamefont {Ceder}},\ }\href {\doibase
  10.1103/PhysRevB.70.235121} {\bibfield  {journal} {\bibinfo  {journal} {Phys.
  Rev. B}\ }\textbf {\bibinfo {volume} {70}},\ \bibinfo {pages} {235121}
  (\bibinfo {year} {2004})},\ \Eprint {http://arxiv.org/abs/0406382}
  {arXiv:0406382 [cond-mat]} \BibitemShut {NoStop}%
\bibitem [{\citenamefont {Hoang}(2014)}]{hoang2014}%
  \BibitemOpen
  \bibfield  {author} {\bibinfo {author} {\bibfnamefont {K.}~\bibnamefont
  {Hoang}},\ }\href {\doibase 10.1039/C4TA04116J} {\bibfield  {journal}
  {\bibinfo  {journal} {J. Mater. Chem. A}\ }\textbf {\bibinfo {volume} {2}},\
  \bibinfo {pages} {18271} (\bibinfo {year} {2014})},\ \Eprint
  {http://arxiv.org/abs/1412.5264} {arXiv:1412.5264} \BibitemShut {NoStop}%
\bibitem [{\citenamefont {Moriya}\ and\ \citenamefont
  {Takahashi}(1984)}]{moriya1984}%
  \BibitemOpen
  \bibfield  {author} {\bibinfo {author} {\bibfnamefont {T.}~\bibnamefont
  {Moriya}}\ and\ \bibinfo {author} {\bibfnamefont {Y.}~\bibnamefont
  {Takahashi}},\ }\href {\doibase 10.1146/annurev.ms.14.080184.000245}
  {\bibfield  {journal} {\bibinfo  {journal} {Annu. Rev. Mater. Sci.}\ }\textbf
  {\bibinfo {volume} {14}},\ \bibinfo {pages} {1} (\bibinfo {year}
  {1984})}\BibitemShut {NoStop}%
\bibitem [{\citenamefont {Massarotti}\ \emph {et~al.}(1997)\citenamefont
  {Massarotti}, \citenamefont {Capsoni}, \citenamefont {Bini}, \citenamefont
  {Chiodelli}, \citenamefont {Azzoni}, \citenamefont {Mozzati},\ and\
  \citenamefont {Paleari}}]{massarotti1997}%
  \BibitemOpen
  \bibfield  {author} {\bibinfo {author} {\bibfnamefont {V.}~\bibnamefont
  {Massarotti}}, \bibinfo {author} {\bibfnamefont {D.}~\bibnamefont {Capsoni}},
  \bibinfo {author} {\bibfnamefont {M.}~\bibnamefont {Bini}}, \bibinfo {author}
  {\bibfnamefont {G.}~\bibnamefont {Chiodelli}}, \bibinfo {author}
  {\bibfnamefont {C.}~\bibnamefont {Azzoni}}, \bibinfo {author} {\bibfnamefont
  {M.}~\bibnamefont {Mozzati}}, \ and\ \bibinfo {author} {\bibfnamefont
  {A.}~\bibnamefont {Paleari}},\ }\href {\doibase 10.1006/jssc.1997.7349}
  {\bibfield  {journal} {\bibinfo  {journal} {J. Solid State Chem.}\ }\textbf
  {\bibinfo {volume} {131}},\ \bibinfo {pages} {94} (\bibinfo {year}
  {1997})}\BibitemShut {NoStop}%
\bibitem [{\citenamefont {N{\'{e}}el}(1948)}]{neel1948}%
  \BibitemOpen
  \bibfield  {author} {\bibinfo {author} {\bibfnamefont {L.}~\bibnamefont
  {N{\'{e}}el}},\ }\href@noop {} {\bibfield  {journal} {\bibinfo  {journal}
  {Ann. Phys. Paris}\ }\textbf {\bibinfo {volume} {3}},\ \bibinfo {pages} {137}
  (\bibinfo {year} {1948})}\BibitemShut {NoStop}%
\end{thebibliography}%

\end{document}